\documentclass[journal]{IEEEtran}

\usepackage{myStyle}
\usepackage{nomencl}
\usepackage{ragged2e}
\usepackage[thinc]{esdiff}
\IEEEoverridecommandlockouts

\newlength{\bracewidth}

\newtheorem{proposition}{Proposition}

\usepackage{etoolbox}
\renewcommand\nomgroup[1]{%
  \item[\bfseries
  \ifstrequal{#1}{P}{A. Parameters}{%
  \ifstrequal{#1}{V}{C. Variables}{%
  \ifstrequal{#1}{S}{B. Sets and Indices}{}}}%
]}

\begin{document}

\title{Scheduling of Software-Defined Microgrids for Optimal Frequency Regulation}

\renewcommand{\theenumi}{\alph{enumi}}

\author{Zhongda~Chu,~\IEEEmembership{Member,~IEEE,} 
        Guoxuan~Cui,~\IEEEmembership{Graduate Student Member,~IEEE,} \\ and
        Fei~Teng,~\IEEEmembership{Senior Member,~IEEE} 
        
        
\vspace{-0.5cm}}
\maketitle
\IEEEpeerreviewmaketitle

\begin{abstract}
Integrated with a high share of Inverter-Based Resources (IBRs), microgrids face increasing complexity of frequency dynamics, especially after unintentional islanding from the maingrid. These IBRs, on the other hand, provide more control flexibility to shape the frequency dynamics of microgrid and together with advanced communication infrastructure offer new opportunities in the future software-defined microgrids. To enhance the frequency stability of microgrids with high IBR penetration, this paper proposes an optimal scheduling framework for software-defined microgrids which aims at combining the control design of the IBR dynamic frequency response and the steady-state economic optimization. This is achieved by utilizing the non-essential load shedding and dynamical optimization of the virtual inertia and virtual damping from IBRs. Moreover, side effects of these services, namely, the time delay associated with non-essential load shedding and potential IBR control parameter update failure are explicitly modeled to avoid underestimations of frequency deviation and over-optimistic results. The effectiveness and significant economic value of the proposed simultaneous and dynamic virtual inertia and damping provision strategy are demonstrated based on case studies in the modified IEEE 33-bus system.
\end{abstract}


\makenomenclature
\renewcommand{\nomname}{\textcolor{black}{Nomenclature}}
\mbox{}
\nomenclature[V]{$H$}{Total microgrid inertia$\,[\mathrm{MWs/Hz}]$}
\nomenclature[V]{$\Delta f$}{Microgrid frequency$\,[\mathrm{Hz}]$}
\nomenclature[P]{$D$}{Total microgrid damping$\,[\mathrm{MW/Hz}]$}

\nomenclature[V]{$\Delta R$}{Primary frequency response from SGs$\,[\mathrm{MW}]$}
\nomenclature[V]{$\Delta P_L$}{Equivalent active power disturbance$\,[\mathrm{MW}]$}
\nomenclature[V]{$\Delta P_{L0}$}{Loss of generation due to islanding events$\,[\mathrm{MW}]$}
\nomenclature[V]{$\Delta P_{L1}$}{Active power disturbance after $T_s\,[\mathrm{MW}]$}
\nomenclature[V]{$\Delta P_{s}$}{Non-essential load shedding$\,[\mathrm{MW}]$}
\nomenclature[P]{$T_s$}{Time delay of non-essential load shedding$\,[\mathrm{s}]$}
\nomenclature[P]{$T_d$}{Fully delivered time of PFR$\,[\mathrm{s}]$}
\nomenclature[V]{$\Delta f_s$}{Frequency deviation due to $T_s\,[\mathrm{s}]$}
\nomenclature[V]{$H_w$}{Virtual inertia from WT $w$ $\,[\mathrm{MWs/Hz}]$}
\nomenclature[V]{$H_b$}{Virtual inertia from storage $b$ $\,[\mathrm{MWs/Hz}]$}
\nomenclature[V]{$H_I$}{Virtual inertia from IBRs $\,[\mathrm{MWs/Hz}]$}

\nomenclature[V]{$\Delta P_w$}{Total additional power supply from WTs $\,[\mathrm{MW}]$}
\nomenclature[V]{$\Delta \Tilde P_{a} $}{Estimated mechanical power deviation $\,[\mathrm{MW}]$}
\nomenclature[P]{$\gamma$}{Auxiliary variable for WT negative damping}
\nomenclature[V]{$\Delta P_b$}{Additional power supply from energy storage system$\,[\mathrm{MW}]$}
\nomenclature[V]{$H_c$}{Total inertia from SGs$\,[\mathrm{MWs/Hz}]$}
\nomenclature[P]{$\mathbf{H}_g$}{Inertia time constant of SGs$\,[\mathrm{s}]$}
\nomenclature[P]{$f_0$}{Nominal frequency$\,[\mathrm{Hz}]$}
\nomenclature[V]{$D_b$}{Virtual damping from storage devices $\,[\mathrm{MW/Hz}]$}
\nomenclature[V]{$D_B$}{Total virtual damping from storage devices $\,[\mathrm{MW/Hz}]$}
\nomenclature[S]{$g\in\mathcal{G}$}{Set of conventional generators}
\nomenclature[S]{$b\in\mathcal{B}$}{Set of storage devices}
\nomenclature[S]{$l\in\mathcal{L}$}{Set of loads}
\nomenclature[S]{$m\in\mathcal{M}$}{Set of PV units}
\nomenclature[S]{$w\in\mathcal{W}$}{Set of wind generation}

\nomenclature[S]{$w\in\mathcal{W}$}{Set of wind generation}

\nomenclature[S]{$\mathcal{H}$}{Set of IBRs providing virtual inertia}
\nomenclature[S]{$\mathcal{D}$}{Set of IBRs providing virtual damping}

\nomenclature[S]{$t\in\mathcal{T}_0$}{Time horizon of frequency events}
\nomenclature[S]{$s\in\mathcal{S}$}{Set of scenarios}
\nomenclature[P]{$\bar P_{b}^{\mathrm{ch}}$/$\bar P_{b}^{\mathrm{dch}}$}{ Maximum charging/discharging rate$\,[\mathrm{MW}]$}
\nomenclature[V]{$P_b$}{Output power of storage devices$\,[\mathrm{MW}]$}
\nomenclature[P]{$\Delta \dot f_\mathrm{lim}$}{Maximum permissible RoCoF$\,[\mathrm{Hz/s}]$}
\nomenclature[P]{$\Delta f_\mathrm{lim}$}{Maximum permissible frequency deviation$\,[\mathrm{Hz}]$}
\nomenclature[P]{$\Delta f_\mathrm{lim}^\mathrm{ss}$}{Maximum permissible steady-state frequency deviation$\,[\mathrm{Hz}]$}
\nomenclature[P]{$\eta_b$}{Efficiency of storage devices}
\nomenclature[P]{$E_{c,b}$}{Energy capacity of storage devices$\,[\mathrm{Hz}]$}
\nomenclature[V]{$\mathrm{SoC}_{b}$}{State of charge of storage devices}
\nomenclature[P]{$P_g^{\mathrm{max}}$}{Capacity of SGs$\,[\mathrm{MW}]$}
\nomenclature[V]{$\Delta f_{\mathrm{lim}}'$}{Equivalent maximum permissible frequency deviation considering noncritical load shedding delay$\,[\mathrm{Hz}]$}

\nomenclature[P]{$P_b^C$}{Capacity of the energy storage device$\,[\mathrm{MW}]$}
\nomenclature[V]{$\bar D_b$}{Maximum virtual damping of the energy storage device$\,[\mathrm{MW/s}]$}

\nomenclature[V]{$x_{1}$}{Auxiliary variable for storage virtual damping constraint reformulation}

\nomenclature[V]{$z_b^H,\,z_b^D$}{Auxiliary variables indicating inertia or damping provision from storage devices}

\nomenclature[P]{$k$}{Robustness degree of the IBR parameter update failure}
\nomenclature[V]{$\mathsf{z}^F$}{Binary vector indicating the IBR parameter update failure}
\nomenclature[V]{$\mathsf{Z}^F$}{Set of all possible $\mathsf{z}^F$}
\nomenclature[V]{$e_{\mathsf k},\,v_{\mathsf k}$}{Auxiliary variables to determine $\mathsf k$ largest virtual inertia}

\nomenclature[P]{$\pi_s$}{Probability associated with scenario s}
\nomenclature[P]{$c_g^{SU}$}{Start up cost of SGs$\,[\mathrm{\pounds}]$}
\nomenclature[P]{$c_g^{R1}$}{Running cost of fixed SGs$\,[\mathrm{\pounds}]$}
\nomenclature[P]{$c_g^{R2}$}{Running cost of flexible SGs$\,[\mathrm{\pounds/MW}]$}
\nomenclature[P]{$c^{VOLL}$}{Value of lost load$\,[\mathrm{\pounds/MW}]$}
\nomenclature[V]{$z_g/y_g$}{Binary variable indicating SG status}
\nomenclature[V]{$p_g$}{Active power produced by SGs$\,[\mathrm{MW}]$}
\nomenclature[V]{$p_l^c/q_l^c$}{Active/reactive load shedding$\,[\mathrm{MW/MVar}]$}
\nomenclature[V]{$p_b^{ch}/p_b^{dch}$}{Charging/discharging power of storage unit b$\,[\mathrm{MW}]$}

\printnomenclature

\section{Introduction}
Microgrids have been an attractive solution to achieve the transition of the electric systems from centralized to decentralized generation for integrating high share of Renewable Energy Sources (RES). In general, microgrids can be viewed as aggregations of loads, Distributed Generators (DGs) and energy storage systems, which operate in coordination to provide reliable electricity to the customers \cite{1354758}. Microgrids are able to operate in both grid-connected and isolated modes. When operating in grid-connected mode, microgrids can exchange power with the maingrid at the Point of Common Coupling (PCC), to achieve maximum economic benefit \cite{6818494}. When the maingrid is subjected to disturbances, microgrids are capable of disconnecting themselves from the maingrid and remain operational with DGs, thus increasing the reliability and resilience of microgrids. 

However, with the continuous development of RES, most DGs are based on solar and wind power. Unlike conventional Synchronous Generators (SGs), these DGs utilize power electronic devices for active power injection, which do not inherently provide inertia or damping to the microgrid. Thus, the microgrid is often regarded as a low inertia and low damping system in which the frequency stability may be challenged by unexpected events or disruptions, especially the unintentional islanding events when the loss of generation from the maingrid can be significant.

{In fact, the dynamic performance of IBRs is programmable through designing the control loop, which potentially offers more support for the microgrid frequency regulation. In conventional microgrids, research efforts are mainly spent on the optimal design of IBR control loop in an offline manner.} The droop control method is commonly applied in microgrids to support frequency regulation after disturbances in a decentralized manner \cite{9625745}. The concept of Virtual Synchronous Generator (VSG) has also been proposed to provide both inertia and damping to the microgrids with high IBR penetration. Based on particle swarm optimization, \cite{9805816} presents a novel parameter tuning method to determine the optimal VSG parameters and virtual impedances from the perspective of small signal stability of islanded microgrids. In \cite{8398456}, an extended VSG is proposed to mimic the inertial and damping response of SGs and a robust control method is utilized to achieve optimal parameter tuning. A VSG based on adaptive virtual inertia control is proposed in \cite{8741094} to further improve the microgrid frequency response through an adaptive fashion. 
Although improvements on microgrid frequency regulation can be achieved based on these static strategies, the increasing variation of system operating conditions makes it difficult to optimally maintain the frequency security by designing a single control strategy without dynamically interacting with other resources through system-level coordination. This also undermines the full advantage of the programmable nature of IBRs.



With the development of advanced  communication systems in microgrids, the IBR control parameters can be dynamically updated based on predefined logic or in response to system level commands \cite{ahmethodzic2021comprehensive}. 
Therefore, it is becoming possible to combine the microgrid system-level scheduling and device-level control design, achieving dynamic optimization of both operation setpoints and IBR control strategies or parameters. {Inspired by software-defined networks that allow fast and accurate reconfiguration of the communication system to provide more flexibility to network operators}, the concept of software-defined microgrids has recently been proposed \cite{9475973}. It refers to the microgrid whose control algorithm, grid structure and operation strategy can be (partially) defined by the implemented software, to realize global optimality in a centralized manner supported by fast and reliable communication. Software-defined microgrids provide more flexibility in system managing and controlling, to improve stability, enhance resilience, and reduce cost. It is suggested in \cite{9475973} that both economics and security of operation should be accounted for in software-defined microgrids, by accommodating the implementation of security and economic features in the microgrid devices. A resilience improvement approach against denial-of-service attacks is proposed in \cite{8525299} by utilizing control plane communication, where the operating strategies of microgrid generators as either voltage sources or current sources are determined by the software-defined control. The authors in \cite{wang2023cyber} propose a novel post-disaster cyber-physical interdependent restoration scheduling framework for active microgrids by coordinating repair crew dispatching, remote-controlled network reconfiguration
and system operation.

Ideally, the IBR frequency control and microgrid operating conditions are desired to be co-optimized in real time to achieve global optimality at any time. However, due to the demanding burden in computation and communication, the real-time co-optimization may not be achievable in practice. In addition, the system operation condition may not have significant changes in a short time interval. Hence, dynamic optimization of IBR control in the same timescale as system scheduling (e.g., half-hourly) is a promising comprise. Reference \cite{9475967} dynamically optimizes the synthetic inertia from IBR in the microgrid scheduling model, whereas the damping provision is not considered. An optimal virtual inertia and damping allocation method is proposed in \cite{shen2023optimal}. {However, the generator commitment decisions which significantly influence the frequency dynamics are taken as input and cannot be co-optimized with the virtual inertia and damping provision. Additionally, the IBR control limits are assumed to be fixed and known, with their dependence on the frequency dynamics being overlooked, thus limiting the overall flexibility and control performance.}

It has also been demonstrated that the non-essential loads in software-defined microgrid can be utilized to support the frequency regulation after an unintentional islanding event. Various load shedding strategies that rely on the frequency and Rate of Change of Frequency (RoCoF) measurements for under-frequency events have been studied and adopted \cite{BAKAR2017161,9770935,7524745,9362316}. Some of the research extends the framework by considering the demand response  \cite{RAFINIA2020102168, 7822924} to guarantee better voltage and frequency stability. In addition, \cite{9475967} proposes a distributionally robust formulation to account for the uncertainty associated with non-essential load shedding. However, load shedding strategies based on frequency measurements and/or real-time communication in software-defined microgrids introduce time delays \cite{dehghanpour2020under}, which is not modeled in these works. This unrealistic assumption may lead to an underestimation of the frequency deviation and jeopardize system security.

{Nevertheless, the research on optimal frequency control and operation in software-defined microgrids by taking full advantage of microgrid communication and IBR control flexibility is limited. Moreover, to the best of our knowledge, no prior work investigates the side-effects due to the utilization of the control flexibility within the software-defined microgrids, namely, the time delay associated with the non-essential load shedding and the potential risk of IBR control parameter update failure. In this context, this paper proposes a software-defined microgrid scheduling model for optimal frequency regulation where the side-effects due to the utilization of the control flexibility within the software-defined microgrids are explicitly modeled and studied.} 
The main contributions of this paper are identified as follows:

\begin{itemize}
    \item {An optimal frequency regulation strategy for software-defined microgrids scheduling is proposed. With simultaneous optimal virtual inertia and damping provision, the proposed software-defined control ensures microgrid frequency security in the most cost-effective manner.}
    \item The frequency security constraints are analytically derived with the time delay of non-essential load shedding due to signal measurement and communication being explicitly modeled within the microgrid frequency dynamics, avoiding underestimations of frequency deviation. 
    \item {Robust frequency constraints against the potential IBR control parameter update failure are formulated for the first time. The factorial increase of frequency constraint number due to the combinatorial nature is further reduced to a linear increase based on the proposed physics-informed reformulation.}
    \item The effectiveness of the proposed approach is validated through case studies in a modified IEEE 33-bus system by comparing with existing state-of-art microgrid frequency regulation strategy. The value and impact of the proposed methodology are fully demonstrated. 
\end{itemize}
The rest of this paper is structured as follows. Section~\ref{sec:2} introduces the microgrid frequency dynamics and derives the frequency metrics. Based on these, frequency constraints are formulated in Section~\ref{sec:3} with the robustness against IBR control parameter update failure considered. Section~\ref{sec:4} describes the overall microgrid scheduling model, followed by case studies in Section~\ref{sec:5} and conclusions in Section~\ref{sec:6}.

\section{Microgrid Frequency Dynamics} \label{sec:2}
The microgrid frequency dynamics after unintentional islanding events is investigated in this section while accounting for the non-essential load shedding and the associated time delay. Meanwhile, the provision of virtual inertia and damping from IBRs are considered with their power and energy limits being explicitly modeled. {
A microgrid containing a set of generation units $g\in \mathcal{G}$ and loads $l\in \mathcal{L}$ is considered here. The generation set is further divided into $\mathcal{G}=\mathcal{G_\mathrm{1}\cup G_\mathrm{2}}$ representing the sets of fast and slow generators. Denote wind, PV and storage units with $w\in \mathcal{W}$, $m\in \mathcal{M}$ and $b\in \mathcal{B}$ respectively.}

\subsection{Frequency Trajectory After Unintentional Islanding Events}
\label{sec:2.1}
\begin{figure}[!t]
    \centering
	\scalebox{1.3}{\includegraphics[]{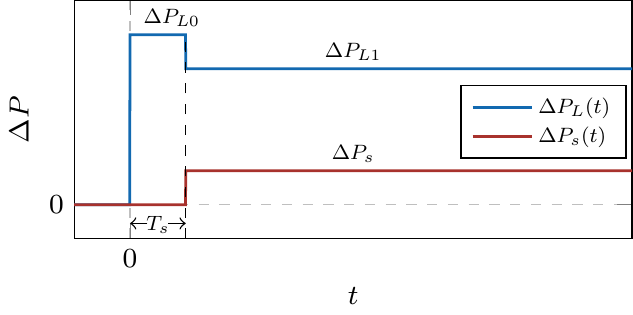}}
    \caption{\label{fig:dP}{{\textcolor{black}{Relationship between disturbance and load shedding.}}}}
    \vspace{-0.35cm}
\end{figure}

The microgrid frequency dynamics can be mathematically described by a single swing equation, under the assumption of the Centre-of-Inertia (CoI) model:
\begin{equation}
    \label{sw1}
    2H\frac{\partial\Delta f(t)}{\partial t} = -D \Delta f(t) + \Delta R(t) -\Delta P_L(t),
\end{equation}
where $\Delta f(t)$ is the microgrid frequency deviation; \textcolor{black}{$H$ and $D$ are the total microgrid inertia and damping respectively, including those from both SGs and IBRs}; $\Delta R(t)$ is the Primary Frequency Responses (PFR)  from conventional SGs; $\Delta P_{L} (t) > 0$ is the equivalent active power disturbance. It can be further expressed as follows:
\begin{equation}
\label{P_L}
    \Delta P_{L} (t) = 
    \begin{cases}
       \Delta P_{L0} &, \; 0 < t\le {T_s} \\ 
       \Delta P_{L1} = \Delta P_{L0} -\Delta P_s &, \; T_s < t
    \end{cases}
\end{equation}
with $\Delta P_{L0}$ being the loss of generation due to the islanding event at $t=0$ and $\Delta P_{s} \in [0,\,\Delta P_{L0}]$ the non-essential load shedding with delay $T_s$. \textcolor{black}{The relationship between the quantities in \eqref{P_L} is visualized in Fig.~\ref{fig:dP}.} It is a common practice to have non-essential load shedding in microgrid after islanding events, in order to ensure the post-event frequency variation within the limits. It can be curtailed in response to unintentional islanding event detections. This process typically involves RoCoF measurements \cite{4523877,KHAMIS2013483} and communication, thus inducing a time delay before the non-essential load is actually shed at $t=T_s$. This time delay if left unaccounted for, may lead to an overestimation of the actual load shedding and insufficiently scheduled frequency services, thus jeopardizing the microgrid frequency stability. Note that both $\Delta P_{L0}$ and $\Delta P_{L1}$ are time-invariant constants, meaning that the equivalent active power disturbance defined in \eqref{P_L} changes stepwise.

Moreover, the PFR $\Delta R(t)$ from conventional SGs can be represented according to the following scheme \cite{6714513}:
\begin{equation}
\label{R}
\Delta R(t)=
     \begin{cases}
       \frac{R}{T_d}t &, \; 0 < t\le  T_d \\ 
       R &, \; T_d< t
     \end{cases}
\end{equation}
with $T_d$ being the PFR delivered time and R being the total PFR delivered by time instant $T_d$. \textcolor{black}{It is assumed that $T_d>T_s$, since the former depending on the grid requirement and the SG time constant, is around 10 s \cite{6714513}, which is larger than the latter which is typically less than 1 s \cite{ten2008evaluation,dehghanpour2020under} due to the RoCoF measurement and communication.} Substituting \eqref{P_L} and \eqref{R} into \eqref{sw1} and solving the resulting differential equation for $0 < t\le T_s$, with the initial condition $\Delta f(0) = 0$, gives the analytical time-domain solution of microgrid frequency during an islanding event:
\begin{equation}
\label{f1(t)}
    \Delta f(t) = \left(\frac{\Delta P_{L0}}{D}+\frac{2HR}{T_d D^2}\right)\left(e^{-\frac{D}{2H}t}-1\right) + \frac{R}{T_d D}t,
\end{equation}
valid $\forall\, t\in (0,\,T_s]$. Similarly, with the initial condition $\Delta f(T_s)$ obtained by evaluating \eqref{f1(t)} at $t=T_s$, the expression of the microgrid frequency deviation for the time period $t\in (T_s,\,T_d]$ can be calculated by:
\begin{align}
\label{f2(t)}
    \Delta f(t) &= \left(\frac{\Delta P_{L1}}{D}+\frac{2HR}{T_d D^2}\right)\left(e^{-\frac{D}{2H}t}-1\right) + \frac{R}{T_d D}t  \nonumber\\
    & + \underbrace{\frac{\Delta P_s}{D} \left(1- e^{\frac{D}{2H}T_s}\right) e^{-\frac{D}{2H}t}}_{\Delta f_s(t)}.
\end{align}
The impact on the frequency trajectory due to the non-essential load shedding delay is reflected through the term $\Delta f_s(t)$. If the time delay $T_s$ is zero, this term disappears and \eqref{f2(t)} becomes the same as \eqref{f1(t)} with $\Delta P_{L0}$ being replaced by $\Delta P_{L1}$, meaning that the size of microgrid disturbance is reduced to $\Delta P_{L1}$ immediately after the unintentional islanding event. Furthermore, this effect exponentially reduces to zero as $t$ approaches infinity, i.e., the delay of non-essential load shedding has no influence on the steady-state frequency, which is also consistent with the intuition. 



\subsection{Modeling of Virtual Inertia from \textcolor{black}{Wind Turbines (WTs)}} \label{sec:2.2}


The control framework proposed in \cite{9066910} is utilized to provide virtual inertia ($H_w$) from WTs based on the kinetic energy extraction. Notably, the virtual damping from WTs is not implemented in this work since the kinetic energy stored in WTs is only capable of supporting overproduction temporarily. 
\textcolor{black}{Furthermore, due to the compromised efficiency of the WT deloading control, it decreases the power output during normal operation, thus being undesired from the perspective of wind farm owners and not considered in this work.}

{In this framework, the additional output power from WTs ($\Delta P_w$) during a system disturbance is given by:
\begin{equation}
    \label{P_w}
    \Delta P_w(t) = - 2 H_{w} \Delta \dot f (t) + \gamma_w H_{w}^2 \Delta f (t),
\end{equation}
\textcolor{black}{with the first term $- 2 H_{w} \Delta \dot f (t)$ being the inertial power and the second term $\gamma_w H_{w}^2 \Delta f (t)$ accounting for the mechanical power deviation due to the WT rotor deceleration. Since the second term is proportional to the frequency deviation, it is equivalent to a negative system damping.} In addition, the total available virtual inertia from WTs in the system (virtual inertia capacity) can be estimated given the wind speed as proposed in \cite{9066910}, where the feasibility of frequency support from WTs and detailed control design can be found as well. It should be noted that, although both $\gamma_w$ and the virtual inertia capacity are operating condition dependent, they can be estimated given the forecast wind speed. In the case of high (above-rated) wind speeds where the pitch control is used for virtual inertia provision, the rotor deceleration and the recovery effect become irrelevant, which can also be represented by setting $\gamma_w=0$.}


\subsection{Modeling of Virtual Inertia and Damping from Energy Storage Devices} \label{sec:2.3}


The additional output power ($\Delta P_b$) of the energy storage device $b\in\mathcal{B}$ after the unintentional islanding event to provide virtual inertia and damping can be expressed by:
\begin{equation}
    \label{dP_b}
    \Delta P_b(t) = - 2 H_{b} \Delta \dot f (t) - D_{b} \Delta f (t),
\end{equation}
where $H_b$ and $D_b$ are the virtual inertia and damping from energy storage unit $b$. {Note that the droop response of IBRs is modeled differently from the SGs' ramp response as defined in \eqref{R} due to the much faster dynamics of IBR control loops.} In order to achieve the optimal and feasible virtual inertia and damping provision from the perspective of microgrid economic operation, the limitation of the storage instantaneous power has to be considered:
\begin{equation}
    \label{P_HD}
    \bar P_{b}^{\mathrm{ch}} \le P_b + \Delta P_b(t) \le \bar P_{b}^{\mathrm{dch}},\,\,\;\forall\, t\in \mathcal{T}_0,
\end{equation}
where $\bar P_{b}^{\mathrm{ch}}$/$\bar P_{b}^{\mathrm{dch}}$ are the maximum charging/discharging rate; $P_b$ is the output during normal operation; $\mathcal{T}_0$ is the time horizon of the frequency event. Power flowing out of the storage unit is defined as the positive direction, i.e., $\bar P_{b}^{\mathrm{ch}}\,<\,0\,<\,\bar P_{b}^{\mathrm{dch}}$. Since only the under-frequency events are focused on, \eqref{P_HD} is rewritten as:
\begin{equation}
    \label{P_HD_max}
    P_b + \max_{t\in \mathcal{T}_0}  \Big\{\left| 2 H_{b} \Delta \dot f (t) + D_{b} \Delta f (t)\right| \Big\} \le \bar P_{b}^{\mathrm{dch}}.
\end{equation}
Although it is possible to derive the maximum absolute value in \eqref{P_HD_max} analytically, the results involve highly nonlinear and nested expressions. Instead, it is typically approximated by the following linear constraints using the triangle inequality \cite{8849022}:
\begin{align}
    \label{P_HD_lim}
    P_b + \max_{t\in \mathcal{T}_0}  \Big\{\left| 2 H_{b} \Delta \dot f (t) + D_{b} \Delta f (t)\right| \Big\} & \le P_b + \nonumber \\
    \max_{t\in \mathcal{T}_0}  \Big\{2 H_{b} \left| \Delta \dot f (t)\right | \Big\} + \max_{t\in \mathcal{T}_0}  \Big\{ D_{b} \left| \Delta f (t)\right| \Big\} & \le P_b + \nonumber \\
    2 H_{b}  \Delta \dot f_{\mathrm{lim}} +  D_{b}  \Delta f_{\mathrm{lim}} \le \bar P_{b}^{\mathrm{dch}},
\end{align}
where $\Delta \dot f_{\mathrm{lim}}>0$ and $\Delta f_{\mathrm{lim}}>0$ are the maximum permissible RoCoF and frequency deviation specified by the system operator. However, due to the fact that the maximum value of RoCoF is attained at $t=0^+$ whereas the frequency nadir is attained at $t_n>0$ as shown in \eqref{tn}, the approximation in \eqref{P_HD_lim} may lead to over-conservative results. Therefore, in order to fully utilize the power capacity of energy storage devices, it is assumed that each energy storage device provides either virtual inertia or virtual damping. This is achieved by introducing binary variables $\{z_b^H,\,z_b^D \,\big | \, z_b^H+z_b^D=1,\, \forall\, b\in \mathcal{B}\}$, which indicate whether the device $b$ provides virtual inertia ($z_b^H$) or damping ($z_b^D$). As a result, the power limit derived in \eqref{P_HD_lim} is rewritten as:
\begin{equation}
    \label{P_lim}
    P_b + 2 H_{b}z_b^H  \Delta \dot f_{\mathrm{lim}} +  D_{b} z_b^D \Delta f_{\mathrm{lim}} \le \bar P_{b}^{\mathrm{dch}},
\end{equation}
where the nonlinear terms ($H_{b}z_b^H$ and $D_{b} z_b^D$) can be effectively linearized by Big-M method. 

As for the energy required by the virtual inertia and damping provision, it should be confined by the following constraint: 
\begin{align}
\label{E_lim}
    \frac{1}{\eta_b}\int_{t\in\mathcal{T}_0} \left| P_b +2 H_{b}z_b^H \Delta \dot f (t) + D_b z_b^D \Delta f(t) \right| dt \nonumber \\ 
    \le  \mathrm{SoC}_b \cdot E_{c,b},
\end{align}
where $\eta_b$, $E_{c,b}$ and $\mathrm{SoC}_b$ are the efficiency, energy capacity and the state of charge of the storage device $b\in\mathcal{B}$. Due to the complex dynamics of $\Delta \dot f (t)$ and $\Delta f(t)$, it is impossible to find the analytical expression of \eqref{E_lim} with a simple structure that can be included in the microgrid operation scheduling. Hence, a piece-wise linear approximation of the RoCoF and frequency trajectory is used instead, to give a conservative result:
\begin{align}
\label{E_D_lim}
    P_b |\mathcal{T}_0|  & + D_b z_b^D \big(\Delta f_{\mathrm{lim}} T_d + \Delta f_{\mathrm{lim}}^\mathrm{ss} \left( |\mathcal{T}_0| - T_d\right) \big) \nonumber \\
    & + H_b z_b^H \Delta \dot f_{\mathrm{lim}} T_d \le  \mathrm{SoC}_b  E_{c,b} \eta_b ,
\end{align}
where $|\mathcal{T}_0|$ covers the time horizon of the entire frequency event. The frequency trajectory is approximated by two segments, with the first being $\Delta f_{\mathrm{lim}}$ from $0$ to $T_d$ and the second $\Delta f_{\mathrm{lim}}^\mathrm{ss}$ (steady-state frequency limit) from $T_d$ to $|\mathcal{T}_0|$. Since the frequency nadir occurs before $T_d$, this approximation is conservative. The RoCoF is approximated by a triangular from $\Delta \dot f_{\mathrm{lim}}$ at $t=0^+$ to $0$ at $t=T_d$. Note that after $|\mathcal{T}_0|$, the restoration process can be conducted by either reconnecting with the utility or generation redispatch, which is out of the scope of this work.

\subsection{Derivation of Frequency Metrics}
In order to deduce the frequency metrics, the frequency support from IBRs as described in the previous sections has to be incorporated by combining \eqref{P_w}-\eqref{dP_b} with \eqref{sw1}. First, the microgrid total inertia $H$ and damping $D$ can be defined as follows:
\begin{equation}
\label{H_total}
    H = H_c + H_I \equiv H_c+ \sum_{w\in \mathcal{W}} H_{w} + \sum_{b\in \mathcal{B}} H_{b} z_b^H,
\end{equation}
which includes the inertia from SGs ($H_c$) and IBRs ($H_I$). $H_c$ takes the form:
\begin{equation}
    \label{Hc}
    H_c =\frac{\sum_{g\in \mathcal{G}} \mathbf{H}_g  P_g^\mathrm{max}}{f_0},
\end{equation}
where $\mathbf{H}_g$ and $P_g^\mathrm{max}$ are the inertia time constant and capacity of SG $g\in \mathcal{G}$; $f_0$ is the system nominal frequency. Similarly, the microgrid damping is formed by:
{
\begin{equation}
\label{D}
    D = D_0 + \underbrace{\sum_{b\in \mathcal{B}} D_{b}z_b^D}_{D_{B}} - \sum_{w\in \mathcal{W}} \gamma_w H_{w}^2,
\end{equation}
with $D_b$ the contribution from energy storage devices.} It should be noted that the microgrid damping is also decreased by $\sum_{w\in \mathcal{W}} \gamma_w H_{w}^2$ due to the virtual inertia provision from WTs \eqref{P_w}. It represents the side effect of virtual inertia provision from wind turbines through kinetic energy extraction, i.e., the output power reduction due to the deviation from the optimal operating point \cite{9066910}.

Based on the time domain solution of the microgrid frequency \eqref{f1(t)} and \eqref{f2(t)}, the analytical expression of the maximum instantaneous RoCoF $(\Delta \dot f_\mathrm{max}\equiv\Delta \dot f|_{t=0^+})$ is identified as:
\begin{equation}
\label{rocof}
    \Delta \dot f|_{t=0^+} = -\frac{\Delta P_{L0}}{2H}.
\end{equation}
Appropriate system inertia $H$ and disturbance size $\Delta P_{L0}$ can be chosen to maintain the maximum RoCoF within the limit. The time instant $t_n$ of frequency nadir is derived by setting the derivative of \eqref{f2(t)} to zero:
\begin{subequations}
\label{tn}
\begin{align}
    t_n &= \frac{2H}{D}\ln{\left(\frac{T_d D \Delta P_{L1}'}{2HR}+1\right)} \\
    \Delta P_{L1}' & = \Delta P_{L0} - \Delta P_{s}e^{\frac{D}{2H}T_s}.
\end{align}
\end{subequations}
Substituting \eqref{tn} into \eqref{f2(t)} leads to the expression for frequency nadir $\left(\Delta f_\mathrm{max}\equiv\Delta f(t_n)\right)$:
\begin{subequations}
\label{f_nadir}
\begin{align}
    \Delta f(t_n) & = \underbrace{\frac{2HR}{T_d D^2} \ln{\left(\frac{T_d D \Delta P_{L1}}{2HR}+1\right)}  -\frac{\Delta P_{L1}}{D}}_{\Delta f_1(t_n)} + \Delta f_s(t_n) \\
    \Delta f_s(t_n) & = \frac{\Delta P_s}{D} \left(1- e^{\frac{D}{2H}T_s}\right) e^{-\frac{D}{2H}t_n}. \label{fs_tn}
\end{align}
\end{subequations}
Note that $\Delta f_s(t_n) \le 0$ is a time-invariant constant, illustrating how the time delay of the non-essential load shedding influences the microgrid frequency nadir. Due to the time delay $T_s$, the frequency nadir is decreased by $\Delta f_s$ compared with the case where the delay is neglected. The steady-state frequency $(\Delta f_\mathrm{max}^\mathrm{ss}\equiv\Delta f(t_\infty))$ can be directly derived by setting $t$ to $\infty$ in \eqref{sw1}:
\begin{equation}
\label{fss}
    \Delta f(t_\infty) = \frac{R - \Delta P_{L_1}}{D}.
\end{equation}



Based on the magnitude of disturbance, both metrics should be kept within prescribed limits by selecting appropriate $H$, $R$ and $D$ terms. However, the highly nonlinear relationship between the frequency nadir and $H$, $D$, $R$, $\Delta P_{L1}$, $\Delta P_{s}$ makes it difficult to incorporate the frequency nadir constraints $|\Delta f(t_n)|\le \Delta f_{\mathrm{lim}}$ into the microgrid scheduling process, which is coped with in next Section. 


\section{Frequency Constraint Formulation}\label{sec:3}
{The frequency nadir constraint derived in the previous section is first converted into \textcolor{black}{Second Order Cone (SOC)} form. Being embedded as operational constraints in the software-defined microgrid scheduling model, these constraints ensure frequency security after unintentional islanding events. In addition, the potential IBR control parameter update failure is considered by formulating the corresponding robust frequency constraints.}

\subsection{Frequency Nadir Constraint}
Based on \eqref{f_nadir}, the frequency nadir constraint $|\Delta f(t_n)| \le \Delta f_\mathrm{lim}$ can be written as:
\begin{equation}
\label{f1_lim}
    |\Delta f_1(t_n)| \le \Delta f_\mathrm{lim}' \equiv \Delta f_\mathrm{lim} -|\Delta f_s(t_n)|
\end{equation}
We achieve the SOC reformulation of the frequency nadir constraint first by utilizing the simplification from \cite{8667397}, the accuracy and validity of which have been fully demonstrated in \cite{8667397,9066910,9475967}. This converts \eqref{f1_lim} into:
    \begin{align}
    \label{nadir_no_Delay}
        HR\ge \frac{\Delta P_{L1}^2T_d}{4\Delta f_\mathrm{lim}'}-\frac{\Delta P_{L1} T_d }{4} D.
    \end{align}
Further combining \eqref{D} and \eqref{nadir_no_Delay} gives the following: {
\begin{align}
\label{nadir}
    HR \ge & \frac{\Delta P_{L1}^2T_d}{4\Delta f_\mathrm{lim}'} - \frac{\Delta P_{L1} T_d (D_0+D_{B})}{4} \nonumber \\
    + & \sum_{w\in \mathcal{W}} \frac{\Delta P_{L1} T_d \gamma_w H_{w}^2}{4}.
\end{align}}
Although \eqref{nadir} resembles a rotated second-order cone, due to the existence of $-\Delta P_{L1} D_{b}$, it cannot be directly incorporated into the optimization model. Hence, an auxiliary variable $x_1$ is introduced to achieve the SOC reformulation of \eqref{nadir}: {
\begin{equation}
\label{x_1}
    x_1^2 = \frac{\Delta P_{L1}^2T_d}{4\Delta f_\mathrm{lim}'} - \frac{\Delta P_{L1} T_d (D_0+D_{B})}{4}.
\end{equation}}
The nonlinear equality constraint \eqref{x_1} can be further enforced by piece-wise linearization \cite{9475967}. It is understandable that due to the complex expression of $|\Delta f_s(t_n)|$ in \eqref{fs_tn}, $\Delta f_\mathrm{lim}'$ is a highly nonlinear term in $D,\,H$ and $t_n$. Hence, it is impossible to incorporate this term directly into the optimization model. Instead, based on historical data, a conservative estimation $[\widehat{D},\,\widehat{H},\,\widehat{t}_n]$ is selected, i.e., $\Delta f_s(t_n) = \left.\Delta f_s(\widehat{t}_n)\right|_{\widehat{D},\widehat{H}}$, which corresponds the largest $|\Delta f_s(t_n)|$ for given $\Delta P_s$ and $T_s$. {Substituting \eqref{x_1} into \eqref{nadir} gives:  
\begin{equation}
    \label{nadir_soc}
    HR \ge x_1^2 + \sum_{w\in \mathcal{W}} \frac{\Delta P_{L1}^{\mathrm{max}} T_d \gamma_w H_{w}^2}{4},
\end{equation}
where $\Delta P_{L1}$ is set to be a constant due to the small magnitude of the coefficient $\gamma_w$. This is an SOC constraint with the decision variables being $H,\,R,\,x_1$ and $H_w$ and can be embedded into the optimization problem directly \cite{chu2023stability1}.}


\subsection{RoCoF and Steady-State Frequency Constraints}
Based on the expressions in \eqref{rocof} and \eqref{fss}, the RoCoF and steady-state frequency constraints can be conveniently formulated as linear constraints as below:
\begin{subequations}
\label{rocof_ss_cstr}
\begin{align}
    2H \Delta \dot f_{\mathrm{lim}}  & \ge \Delta P_{L0}\\
    R+D \Delta f_{\mathrm{lim}}^\mathrm{ss} & \ge \Delta P_{L1},
\end{align}
\end{subequations}
with $\Delta \dot f_{\mathrm{lim}}$ and $\Delta f_{\mathrm{lim}}^\mathrm{ss}$ being the maximum permissible RoCoF and steady-state frequency deviation respectively.

\subsection{Robust Frequency Constraints Against Parameter Update Failure} \label{sec:3.3}
\color{black}
The dynamic optimization of IBR control parameters requires frequent updates and modifications of the IBR control loop in response to the command from the control center based on communication in software-defined microgrids. However, it also brings risks in frequency maintenance when certain communication signals are not captured by the desired unit due to bandwidth limitations, environment disturbances or even cyber-attacks. In order to account for these situations, the frequency constraints have to be robust against potential IBR control parameter update failure. \textcolor{black}{Here, it is assumed that when an IBR unit does not receive the signal of virtual inertia or damping from control center, it does not provide any service to the microgrid. Note that although it is possible to assume that the virtual inertia or damping of last time step is provided, this may cause IBR overloading and frequency constraint violation due to the operating point variation, thus not being considered.} In this work, the robust frequency constraints against update failure of any $k$ control parameters (virtual inertia or damping) is considered, where $k\le|\mathcal{I}|=|\mathcal{W}\,\cup\, \mathcal{B}|$ is a constant specified by system operators. 

\textcolor{black}{Ideally, it is desired to directly identify the worst case in terms of the microgrid frequency response from all the possible $C_k^{|\mathcal{I}|}$ combinations. Intuitively, this worst case corresponds to removing $k$ frequency responses from IBRs (virtual inertia/damping) with the largest contributions. However, it is challenging to determine the compositions of this $k$ frequency responses, since the effect of inertia and damping on support the frequency evolution varies with the operating conditions and the values of other frequency dynamics-related variables and parameters. To overcome this challenge, we} first define a vector containing binary parameters with dimension $|\mathcal{I}|$: $\mathsf{z}^F\in\{0,1\}^{|\mathcal{I}|}$, where each element in $\mathsf{z}^F$ represents whether the control parameter in $i\in \mathcal{I}$ fails to be updated ($\mathsf{z}^F_i = 0$) or not ($\mathsf{z}^F_i = 1$). Furthermore, $H_I$ in \eqref{H_total} and $D_B$ in \eqref{D} are replaced by:
\begin{subequations}
\label{HD_1}
\begin{align}
    H_I^F & = \sum_{w\in \mathcal{W}} H_{w} \mathsf{z}^F_w+\sum_{b\in \mathcal{B}} H_{b} z_b^H \mathsf{z}^F_b \\
    D_B^F & = \sum_{b\in \mathcal{B}} D_{b} z_b^D \mathsf{z}^F_b,
\end{align}
\end{subequations}
where $\mathsf{z}^F_w$ and $\mathsf{z}^F_b$ are elements in $\mathsf{z}^F$. With the above relationship, the robust frequency constraints can be expressed by:
\begin{align}
\label{freq_combination}
    &\eqref{nadir_soc},\,\eqref{rocof_ss_cstr}, & \forall\, &\mathsf{z}^F \in \mathsf{Z}^F.
\end{align}
$\mathsf{Z}^F$ is the set of all possible $\mathsf{z}^F$ in which the number of zero element equals $k$:
\begin{equation}
\label{Z^F}
    \mathsf{Z}^F = \left\{\mathsf{z}^F\in\{0,1\}^{|\mathcal{I}|}\,\big | \, \mathbf{1} ^{\mathsf{T}} \mathsf{z}^F = |\mathcal{I}| - k   \right\},
\end{equation}
with $\mathbf{1}$ denoting a vector of ones with a comfortable dimension. Equation \eqref{freq_combination} requires the frequency constraints to be held for all possible situations that satisfy \eqref{Z^F}, which is essentially combinatorial optimization. As $k$ and $|\mathcal{I}|$ grow, the number of frequency constraints increases significantly due to the combinatorial explosion ($C^{|\mathcal{I}|}_k$), which limits its large-scale application. The challenge lies in identifying the worst case among all the possible combinations in $\mathsf{Z}^F$, due to the complex dependence between inertia, damping and frequency dynamics, i.e., the loss of frequency support from which $k$ units in $\mathcal{I}$ would lead to the worst frequency dynamics. In order to overcome this issue, a physics-informed reformulation is proposed to reduce the number of frequency constraints considerably. First, classify all the combinations in \eqref{Z^F} into $k+1$ groups ($\mathsf{Z}^F_{\mathsf{k}}$), according to the number of virtual inertia and damping failure, i.e., $\forall\, \mathsf{k}\in\Set{0,1, ..., k}$,
\begin{equation}
    \mathsf{Z}^F_{\mathsf{k}} = \Set{\mathsf{z}^F \in\mathsf{Z}^F\ | \begin{array}{l}
    \sum_{i\in\mathcal{H}} (1-\mathsf{z}^F_i) = \mathsf{k} \\
    \sum_{i\in\mathcal{D}} (1-\mathsf{z}^F_i) = k-\mathsf{k}
  \end{array} },
\end{equation}
which represents the situation where $\mathsf{k}$ units fail to provide virtual inertia and $k-\mathsf{k}$ units fail to provide virtual damping. $\mathcal{H}$ and $\mathcal{D}$ are the sets of IBRs that provide virtual inertia and damping respectively:
\begin{subequations}
\begin{align}
    \mathcal{H} &= \mathcal{W} \cup  \mathcal{B}_H, & \mathcal{B}_H &= \Set{b\in \mathcal{B}\ | z_b^H = 1,\,z_b^D = 0} \\
    \mathcal{D} &=  \mathcal{B}_D, & \mathcal{B}_D &=\Set{b\in \mathcal{B}\ | z_b^D = 1,\,z_b^H = 0}
\end{align}
\end{subequations}
Although the set $\mathsf{Z}^F$ is divided into $k+1$ groups, the total number of all the combinations has not been reduced, i.e., $\sum_{\mathsf{k}} |\mathsf{Z}^F_{\mathsf{k}}| =  |\mathsf{Z}^F|$. Next, it is demonstrated by Proposition~\ref{Prop_1} that the worst case in each $\mathsf{Z}^F_{\mathsf{k}}$ can be identified offline, which is further constrained by a single set of frequency constraints (RoCoF, nadir and steady-state). 

\begin{proposition} \label{Prop_1}
$\forall\, \mathsf{k}\in\Set{0,1, ..., k}$, $\exists \, \mathsf{z}^{F,*}_{\mathsf{k}} \in \mathsf{Z}^F_{\mathsf{k}}$, such that $\left| \Delta f_{\mathrm{max}}(\mathsf{z}^{F,*}_{\mathsf{k}}) \right | \ge \left| \Delta f_{\mathrm{max}}(\mathsf{z}^{F}) \right |$, $\forall\, \mathsf{z}^{F} \in \mathsf{Z}^F_{\mathsf{k}}$, where $\mathsf{z}^{F,*}_{\mathsf{k}}  \in \mathsf{Z}^F_{\mathsf{k}}$ satisfies:
\begin{subequations}
\begin{align}
    \sum_{i\in\mathcal{H}} (1- \mathsf{z}^{F,*}_{\mathsf{k},i}) H_i &\ge \sum_{i\in\mathcal{H}} (1- \mathsf{z}^F_i) H_i, & &\forall\, \mathsf{z}^F \in \mathsf{Z}^F_{\mathsf{k}}  \\
    \sum_{i\in\mathcal{D}} (1- \mathsf{z}^{F,*}_{\mathsf{k},i}) D_i &\ge \sum_{i\in\mathcal{D}} (1- \mathsf{z}^F_i) D_i, & &\forall\, \mathsf{z}^F \in \mathsf{Z}^F_{\mathsf{k}}.
\end{align}
\end{subequations}
\end{proposition}
Proposition~\ref{Prop_1} can be proved based on the fact, ${\partial |\Delta f_{\mathrm{max}}|}/{\partial H}<0$ 
and ${\partial |\Delta f_{\mathrm{max}}|}/{\partial D}<0$, which is not detailed here. It holds for $\Delta f_{\mathrm{max}}^{\mathrm{ss}}$ and $\Delta \dot f_{\mathrm{max}}$ as well. As a result, \eqref{HD_1} can be transformed into the following, to represent the worst case corresponding to $\mathsf{z}^{F,*}_{\mathsf{k}}$, $\forall\, \mathsf{k}\in\Set{0,1, ..., k}$:
\begin{subequations}
\label{HD_2}
\begin{align}
    H_{I,\mathsf{k}}^F & = H_I - H_{\mathsf{k}},  \\
    D_{B,\mathsf{k}}^F & = D_B - D_{\mathsf{k}},
\end{align}
\end{subequations}
where $H_{\mathsf{k}}$ and $D_{\mathsf{k}}$ are the sum of $\mathsf{k}$ largest virtual inertia and $k-\mathsf{k}$ largest virtual damping respectively. This can be further enforced by the following constraints, by introducing a scalar $e_{\mathsf{k}}\in \mathbb{R}$ and a vector $v_{\mathsf{k}}\in \mathbb{R}^{|\mathcal{I}|}$ , $\forall\, \mathsf{k}\in\Set{0,1, ..., k}$ \cite{ogryczak2003minimizing}.
\begin{subequations}
\label{k_largest}
\begin{align}
    H_{\mathsf{k}} &\ge  \mathsf{k} e_{\mathsf{k}} + \sum_{i\in \mathcal{I}} v_{\mathsf{k}, i}\\
    v_{\mathsf{k}, i} &\ge 0, & &\forall\, i\in \mathcal{I}\\
    e_{\mathsf{k}} + v_{\mathsf{k}, w} &\ge H_w, & &\forall\, w\in \mathcal{W}\\
    e_{\mathsf{k}} + v_{\mathsf{k}, b} &\ge H_b z^H_b, & &\forall\, b\in \mathcal{B}
\end{align}
\end{subequations}
$D_{\mathsf{k}}$ can be constrained in a similar fashion, which is not covered here. With \eqref{HD_2} and \eqref{k_largest}, the sum of $\mathsf{k}$ largest virtual inertia and $k-\mathsf{k}$ largest damping from IBRs will be reduced from the total values in the optimization, leading to the following robust frequency constraints:
\begin{align}
\label{freq_robust}
    &\eqref{nadir_soc},\,\eqref{rocof_ss_cstr}, & \forall\, &\mathsf{k}\in\Set{0,1, ..., k},
\end{align}
which reduces the total number of frequency constraints from $3\times C_k^{|\mathcal{I}|}$ in \eqref{freq_combination} to $3\times (k+1)$. 


\section{Software-Defined Microgrid Scheduling with Optimal Frequency Regulation} \label{sec:4}
\begin{figure}[!t]
    \centering
	\scalebox{0.39}{\includegraphics[]{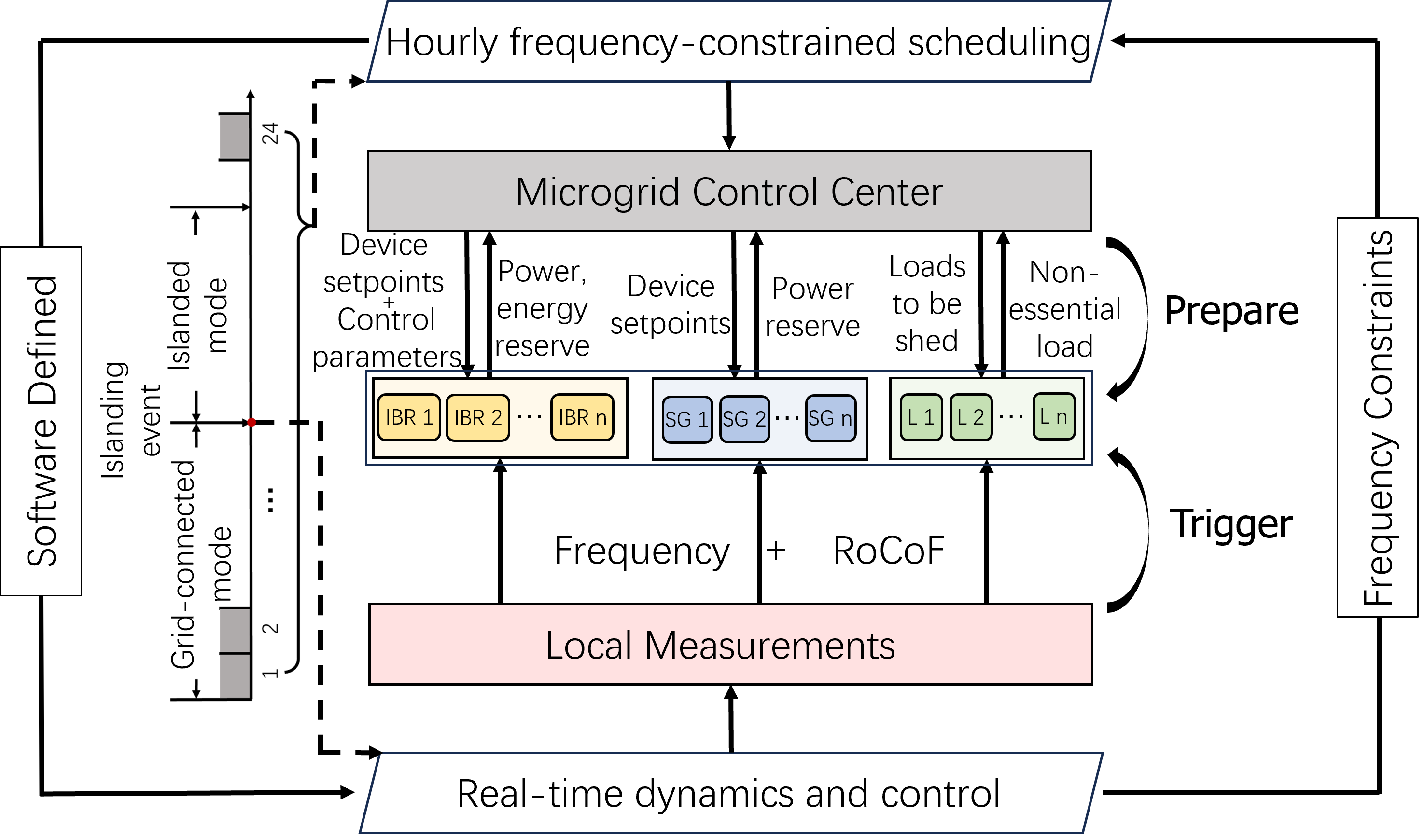}}
    \caption{\label{fig:timescale}{{\textcolor{black}{Overall framework of the proposed software-defined microgrid scheduling.}}}}
    \vspace{-0.35cm}
\end{figure} 

The proposed method of simultaneous provision of virtual inertia and damping for optimal frequency regulation is to be implemented into the software-defined microgrid scheduling model, which is responsible for determining the optimal generator commitment and dispatch, wind/PV generation and curtailment, charge/discharge power and state of charge of storage devices, load shedding as well as the optimal frequency services (i.e., PFR from SGs, virtual inertia and damping from WTs and storage units). {A microgrid scheduling model based on the two-stage stochastic optimization} is introduced in this section to demonstrate the connection between frequency regulation and microgrid scheduling. \textcolor{black}{The overall architecture of the proposed method is demonstrated in Fig.~\ref{fig:timescale}. During normal operation, the microgrid operates according to the results of hourly frequency-constrained scheduling. With the optimal frequency services and setpoints being co-optimized and updated in each hour, the microgrid can ensure frequency security cost-effectively. The frequency services are triggered only during the transition from normal operation to islanded operation. Hence, updating the virtual inertia and damping of IBRs during normal operation has little impact on their overall performances and stability. The proposed software-defined framework lies in the nature of dynamically optimizing and updating the IBR control parameters for optimal frequency regulation.}

In order to manage the uncertainties associated with renewable generation and demand in the microgrid, the two-stage decision process is introduced. In the first stage, the unit commitment decisions are made for the slow generators whereas in the second stage, the power generation of committed units as well as the fast-start generators are decided to meet the load, once most of the uncertain inputs (demand and renewable generation) are realized \cite{Conejo_2010}.

\subsection{Objective Function} \label{sec:4.1}
The objective of the scheduling problem is to minimize the microgrid average operation cost for all scenarios ($\forall\, s \in \mathcal{S}$) along with considered time horizon $t\in \{0,1,...,T\}$:
\begin{equation}
\begin{split}
    \min & \sum_{s\in \mathcal{S}} \sum_{t\in T} \pi_s(\sum_{g\in \mathcal{G}}c_g^{SU}z_{t,s,g}+\Delta t(\sum_{g\in \mathcal{G_\mathrm{1}}}c_g^{R1}y_{t,s,g}+ \\ & \sum_{g\in \mathcal{G_\mathrm{2}}}c_g^{R2}p_{t,s,g}
     +\sum_{l\in \mathcal{L}}c^{VOLL}(p_{t,s,l}^{c}+(q_{t,s,l}^{c})^2)))
\end{split}
\end{equation}
where $\pi_s$ is the probability associated with scenario $s$; $c_g^{SU}$, $c_g^{R1}/c_g^{R2}$ and $c^{VOLL}$ refer to start-up costs, running costs of fixed/flexible generators and the value of lost load (VOLL); $z_{t,s,g}$ and $y_{t,s,g}$ are binary variables of generator $g$ at time step $t$ in scenario $s$ with $1/0$ indicating starting up/not and being on/off; $p_{t,s,g}$ and $p_{t,s,l}^{c}/q_{t,s,l}^{c}$ denote the active power produced by generators and active/reactive load shedding. \textcolor{black}{The scenario tree is built based on user-defined quantiles of the forecasting error distribution to capture the uncertainty associated with the demand and wind generation. Two steps are involved to generate the scenarios: (i) creating the distribution of net demand;  (ii) quantifying the value of net demand and its probability in each node. The readers are referred to \cite{7115982} for more information.}

\subsection{Constraints} \label{sec:4.2}
The conventional microgrid scheduling constraints related to the generator operation, power flow limits, power balance, wind/PV curtailment and load shedding are not included here. \cite{9475967} can be referred to for more details. 
\subsubsection{Constraints of battery storage system}
\begin{subequations}
\begin{align}
    \label{p_b}
    & \eqref{P_lim}\, \eqref{E_D_lim}, \;\;\;\;\; \forall\, t,s,b\\
    &\mathrm{SoC}_{t,s,b}E_{c,b} = \mathrm{SoC}_{t-1,s,b}E_{c,b} - \frac{1}{\eta_b} p_{t,s,b}^{\mathrm{dch}}\Delta t \nonumber \\
    \label{soc_cal}
    & + {\eta_b} p_{t,s,b}^{\mathrm{ch}}\Delta t, \;\;\;\;\; \forall\, t,s,b \\
    & p_{t,s,b} = p_{t,s,b}^{\mathrm{dch}} - p_{t,s,b}^{\mathrm{ch}}, \;\;\;\;\; \forall\, t,s,b\\
    \label{soc_lim}
    & \mathrm{SoC}_{\mathrm{min}} \le \mathrm{SoC}_{t,s,b} \le \mathrm{SoC}_{\mathrm{max}}, \;\;\;\;\; \forall\, t,s,b\\
    \label{SoC_T}
    & \mathrm{SoC}_{0,s,b} = \mathrm{SoC}_{T,s,b}, \;\;\;\;\; \forall\, t,s,b.
\end{align}
\end{subequations}
The power injection from the battery storage system to the microgrid is confined in \eqref{p_b} by the upper bound of the charging and discharging rate with $p_{b}$ in the original equations being replaced by $p_{t,s,b}$. The battery state of charge is quantified by \eqref{soc_cal} with the charging/discharging efficiency $\eta_b$. \eqref{soc_lim} imposes the upper and lower limits on the SoC of the storage devices. The SoC at the end of the considered time horizon is set to be a pre-specified value being equal to its initial value as in \eqref{SoC_T}.

\subsubsection{Frequency security constraints subsequent to islanding events}
According to the derivation in Section \ref{sec:3}, \eqref{nadir_soc}, \eqref{rocof_ss_cstr} are incorporated into the microgrid scheduling model as the frequency nadir, RoCoF and steady-state constraints, or \eqref{freq_robust} if the robustness against parameter update failure is considered.  Therefore, the optimal microgrid inertia which includes that from both conventional SGs and IBRs, the virtual damping, the PFR from SGs and the imported power from the maingrid will be determined in the microgrid scheduling model to ensure the minimum operational cost while maintaining the frequency constraints.

\subsubsection{Constraints of IBR control parameter update frequency} \label{sec:4.2.3}
Due to the efforts on control parameter modification, which may incur additional operation and maintenance costs, IBR owners may be reluctant to change IBR control parameters too frequently. On the other hand, from the perspective of microgrid operators, it may not be necessary to change these parameters at each time step in the scheduling, especially considering the potential risk due to the cyber-security concern. Therefore, in order to enforce a maximum number ($\tau_{\mathrm{max}}$) of the IBR control parameter changes within one-day operation, the following constraints can be included in the proposed software-defined microgrid scheduling framework. 
\begin{equation}
\label{tau_max}
    \sum_{t =1}^T \tau_t \le \tau_{\mathrm{max}},
\end{equation}
where $\tau_t,\, \forall t \in \Set{1,2,..., T}$ are binary variables indicating if the IBR control parameters change at time step $t$, which can be formally defined as below. 
\begin{align}
    \label{tau_t}
    \tau_t &= \left\{\begin{array}{@{}cl}
                0, & \mathrm{if}\, \begin{array}{lc}
                     H_{w,t} = H_{w,t-1},\, \forall w \in \mathcal{W}, \,\wedge \\
                     H_{b,t} z_{b,t}^H  = H_{b,t-1} z_{b,t-1}^H,\, \forall b \in \mathcal{B}, \,\wedge \\
                     D_{b,t} z_{b,t}^D  = D_{b,t-1} z_{b,t-1}^D,\, \forall b \in \mathcal{B}
                \end{array} \\
                1,   & \mathrm{otherwise}\\
                \end{array} \right.
\end{align}
The conditional constraints in \eqref{tau_t} can be conveniently converted to linear form using Big-M method, which is not discussed here. 
\color{black}

\section{Case studies} \label{sec:5}
In order to demonstrate the performance of the proposed distributionally robust chance-constrained microgrid scheduling model, case studies are carried out through the modified IEEE 33-bus distribution system \cite{9258930}.
{The optimization problem is solved in a horizon of 24 hours with the time step being 1 hour.} System parameters are set as follows: load demand $P_D\in [3.16,5.19]\,\mathrm{MW}$, load damping $D_0 = 0.5\% P^D / 1\,\mathrm{Hz}$, PFR delivery time $T_d = 10\,\mathrm{s}$. The frequency limits of nadir, steady-state value and RoCoF are set as: $\Delta f_\mathrm{lim} = 0.8\,\mathrm{Hz}$, $\Delta f_\mathrm{lim}^\mathrm{ss} = 0.5\,\mathrm{Hz}$ and $\Delta \dot f_\mathrm{lim} = 1\,\mathrm{Hz/s}$. The time delay of non-essential load shedding is $T_s = 0.4\,\mathrm{s}$, except in Section~\ref{sec:5.2} where the impact of this time delay is investigated.

\begin{figure}[!b]
    \centering
	\scalebox{0.26}{\includegraphics[]{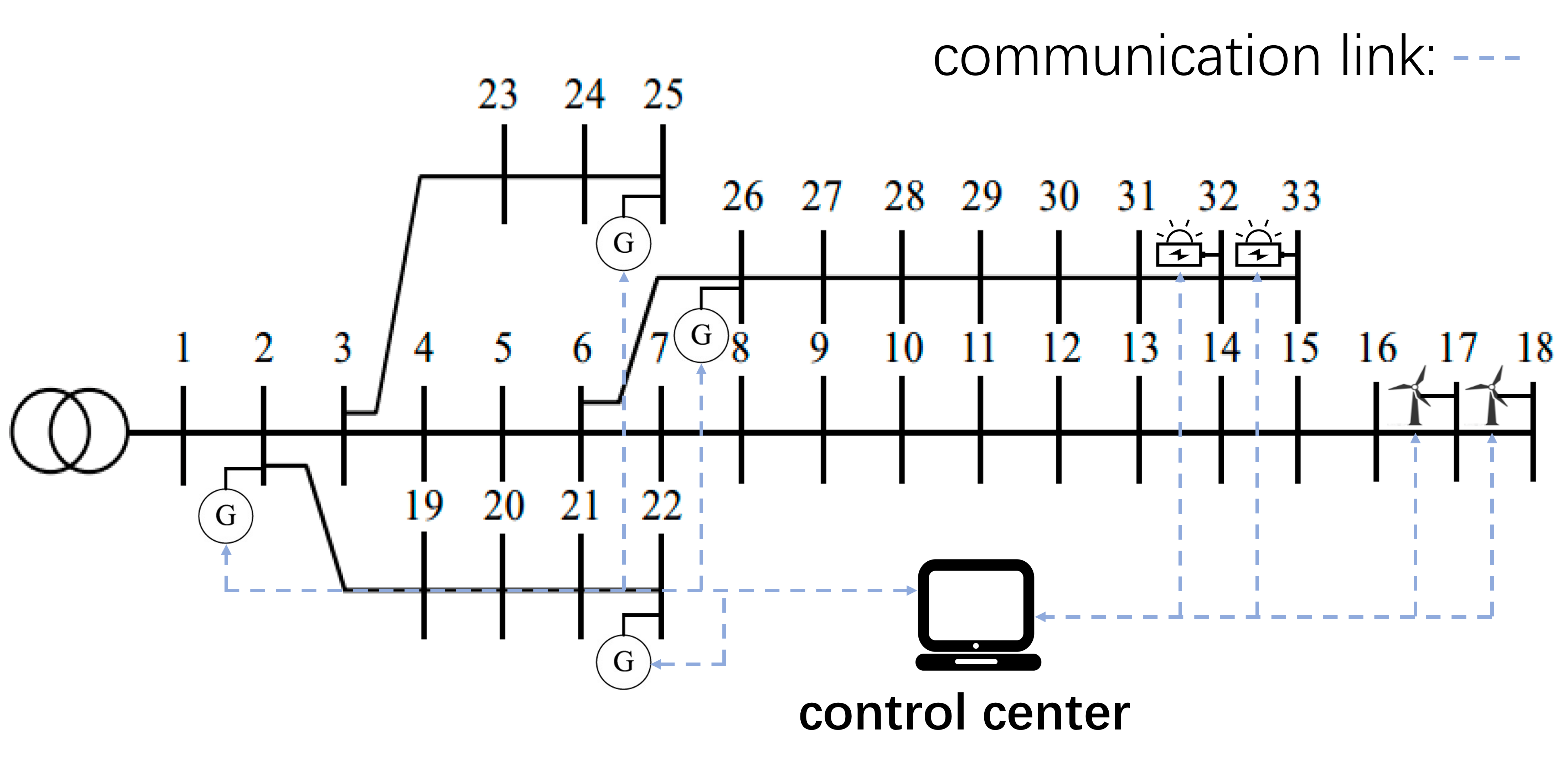}}
    \caption{\label{fig:33_bus}{\textcolor{black}{Single line diagram of modified IEEE 33-bus system.}}}
\end{figure}

Dispatchable SGs are installed at Bus 2 22 25 and 26 with a total capacity of $4.5\,\mathrm{MW}$. The PV-storage system and wind turbines are located at Bus 32,33 and 17,18, with total capacities of $2\,\mathrm{MW}$ and $1.2\,\mathrm{MW}$ respectively, as shown in Fig.~\ref{fig:33_bus}. 
\color{black}
For synchronous generators, we consider a traditional model equipped with a prime mover and a \textit{TGOV1} governor. Furthermore, the automatic voltage regulator based on a simplified excitation system \textit{SEXS} is incorporated, together with a \textit{PSS1A} power system stabilizer~\cite{entsoeGen}. Internal machine dynamics are characterized by the transients in the rotor circuits described through flux linkage. For more details regarding the generator modeling and internal parameter computation, we refer the reader to~\cite{Kundur1994}.

For IBR dynamics, we consider a state-of-the-art VSC control scheme previously described in \cite{8579100,UrosISGTeurope}, where the outer control loop consists of active and reactive power controllers providing the output voltage angle and magnitude reference by adjusting the predefined setpoints according to a measured power imbalance. Subsequently, the reference voltage vector signal is passed through a virtual impedance block, as well as the inner control loop consisting of cascaded voltage and current controllers. The output is combined with the DC-side voltage in order to generate the modulation signal. More details on the overall converter control structure and employed parametrization can be found in \cite{8579100,UrosISGTeurope}.
\color{black}

The aggregated parameters of battery devices at each bus are listed in Table~\ref{tab:battery}. The weather conditions are obtained from online numerical weather prediction \cite{weather}. The MISOCP-base optimization problem is solved by Gurobi (10.0.0) on a PC with Intel(R) Core(TM) i7-7820X CPU @ 3.60GHz and RAM of 64 GB. 


\begin{table}[!t]
\vspace{-0.35cm}
\renewcommand{\arraystretch}{1.2}
\caption{Parameters of Battery Storage Devices}
\label{tab:battery}
\noindent
\centering
\vspace{-0.25cm}
    \begin{minipage}{\linewidth} 
    \renewcommand\footnoterule{\vspace*{-5pt}} 
    \begin{center}
        \begin{tabular}{ c | c | c | c | c }
            \toprule
             $\boldsymbol{\mathrm{SoC}_{\mathrm{min}}}$ &$\boldsymbol{\mathrm{SoC}_{\mathrm{max}}}$ &$\boldsymbol{\eta_b}$ &$\boldsymbol{|\bar P^{\mathrm{(d)ch}}|\,\mathrm{[MW]}}$  & $\boldsymbol{E_c\,\mathrm{[MWh]}}$ \\ 
            \cline{1-5}
            $15\%$ & $85\%$  & $0.9$  & $0.5$ & $1.5$ \\ 
           \bottomrule
        \end{tabular}
        \end{center}
    \end{minipage}
    \vspace{-0.35cm}
\end{table} 

\subsection{Validation of Proposed Frequency Constraints}\label{sec:5.1}
\begin{figure}[!b]
    \centering
    \vspace{-0.35cm}
	\scalebox{1.2}{\includegraphics[]{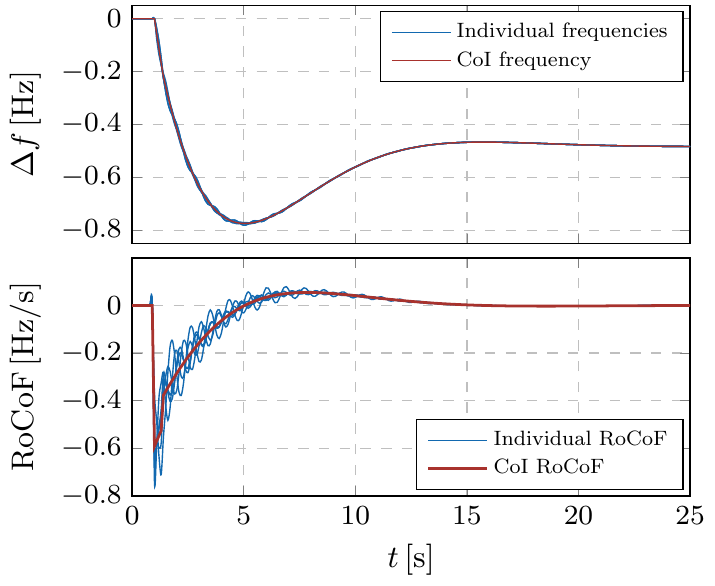}}
    \caption{\label{fig:f}{\textcolor{black}{Microgrid frequency and RoCoF evolution after an islanding event.}}}
    \vspace{-0.35cm}
\end{figure}
To validate the accuracy of the proposed frequency dynamics and frequency constraints, sample solutions of the microgrid scheduling with the proposed optimal frequency regulation framework are fed to the dynamic simulation model implemented in Matlab/Simulink. The resulting evolution of microgrid frequency and RoCoF are depicted in Fig.~\ref{fig:f}, with the CoI frequency nadir of $0.774\,\mathrm{Hz}$ and maximum RoCoF of $-0.61\,\mathrm{Hz/s}$ being within prescribed limits. The dramatic increase of the RoCoF from $-0.52\,\mathrm{Hz/s}$ to $-0.38\,\mathrm{Hz/s}$ at $t = T_s$ is due to the non-essential load shedding.

\begin{figure}[!t]
    \centering
    \vspace{-0.35cm}
	\scalebox{1.2}{\includegraphics[]{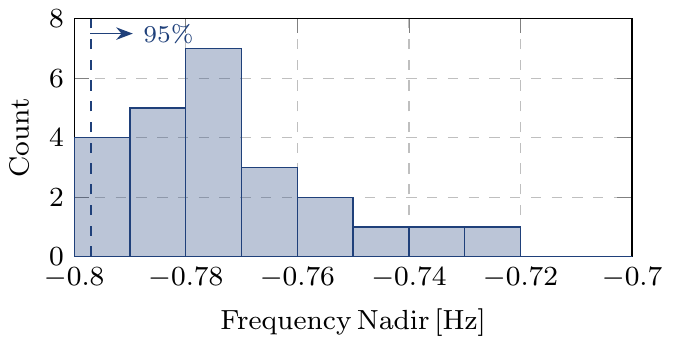}}
    \caption{\label{fig:Hitogram}{{Frequency nadir histogram with 95\% confidence interval of one-day scheduling.}}}
    \vspace{-0.35cm}
\end{figure}

All the other trajectories present similar trends in frequency and RoCoF, thus not being covered here. Instead, to demonstrate the robustness of the proposed method in terms of the effectiveness of the nadir constraints, the frequency nadir in each hour of the one-day scheduling, if an unintentional islanding event occurs, is obtained through the dynamic simulation with the results depicted in Fig.~\ref{fig:Hitogram}. It is observed from the histogram that all the frequency nadirs during the 24-hour scheduling are close to the boundary ($-0.8\,\mathrm{Hz}$) with the mean and standard deviation being $-0.7728\,\mathrm{Hz}$ and $0.018\,\mathrm{Hz}$ respectively, indicating good conservativeness and robustness of the proposed approximation. {The 95\% confidence interval is also shown in the figure by the blue dashed line, i.e., $\mathbf{Pr}\{\Delta f_{\mathrm{max}}\ge -0.797\}\ge 0.95$.} 


{Furthermore, the values of [$\hat{D}$, $\hat{H}$, $\hat{t}_n$] is [$0.778\,\mathrm{MW/Hz}$, $1.229\,\mathrm{MWs/Hz}$, $3.292\,\mathrm{s}$], which is selected from the range: $[0.726,0.994]\,\mathrm{MW/Hz}$, $[1.033, 1.404]\,\mathrm{MWs/Hz}$, $[3.292,4.874]\,\mathrm{s}$ respectively. In this case, $\left.\Delta f_s(\widehat{t}_n)\right|_{\widehat{D},\widehat{H}}$ takes the value of $-0.0314\,\mathrm{Hz}$, showing a conservative yet acceptable approximation, compared with the real values $\Delta f_s(t_n)\in[-0.0314,-0.0133]$. Nevertheless, this approximation does not influence the system operation and the optimal solution significantly, as the impact of the time delay on the magnitude of the frequency  nadir is small, as demonstrated in Section~\ref{sec:5.3}.}

\subsection{Value of Simultaneous Virtual Inertia and Damping Provision} \label{sec:5.2}
The value of simultaneous provision of virtual inertia and damping from IBRs is assessed in this subsection with four different cases defined as, Case I/II: without/with virtual inertia, without virtual damping; Case III/IV: without/with virtual inertia, with virtual damping.


\begin{figure}[!b]
    \centering
    \vspace{-0.35cm}
	\scalebox{1.2}{\includegraphics[]{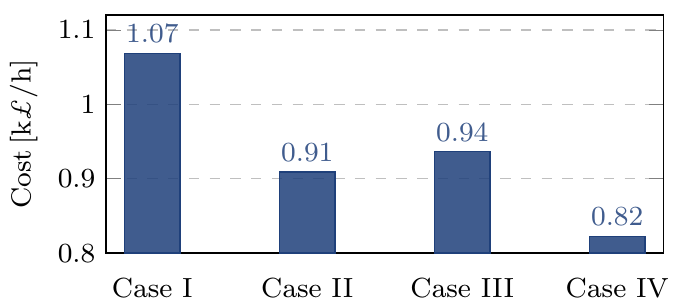}}
    \caption{\label{fig:Cost_damping}{Microgrid operation cost with different frequency services.}}
\end{figure}

The averaged microgrid operation cost is plotted in Fig.~\ref{fig:Cost_damping}. As expected, Case I has the highest cost ($1.07\,\mathrm{k\pounds /h}$) since no virtual inertia or damping is provided from IBRs to support the frequency dynamics after an unintentional islanding event. To ensure the frequency security constraints in this case, more SGs are dispatched online to provide inertia and frequency response and less power can be utilized from IBR. On the contrary, if virtual inertia is allowed from IBRs (Case II), the operation cost reduces to $0.91\,\mathrm{k\pounds /h}$ as fewer SGs are needed and more renewable energy can be utilized. A similar result is observed for Case III, where the operation cost is somewhat higher than that of Case II, meaning that solely providing virtual damping is slightly less effective on frequency regulation compared with inertia provision only. 

Furthermore, with the proposed method where the optimal virtual inertia and damping can be simultaneously dispatched during the microgrid scheduling (Case IV), the microgrid operation cost can be decreased by about 10\% to $0.82\,\mathrm{k\pounds /h}$ even compared with Case II, which demonstrates the significant value of optimizing virtual inertia and damping provision from IBRs together. 

\begin{figure}[!t]
    \centering
    \vspace{-0.35cm}
	\scalebox{1.2}{\includegraphics[]{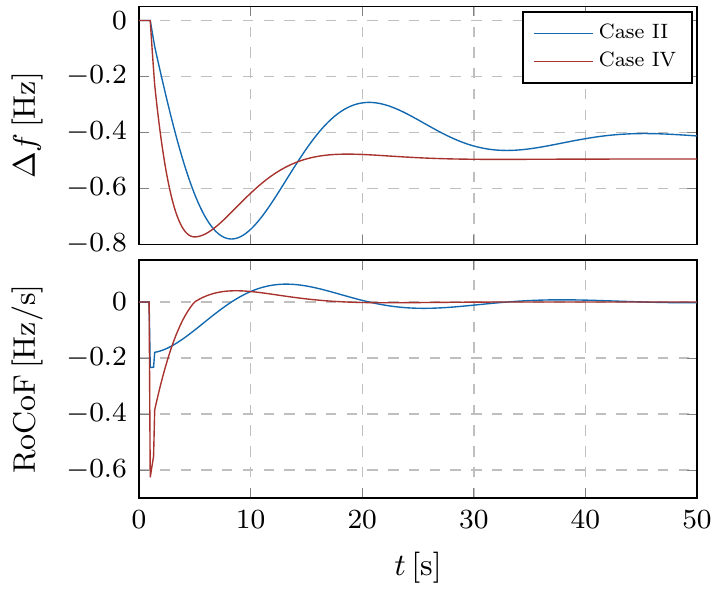}}
    \caption{\label{fig:f_HD}{Microgrid frequency and RoCoF trajectories with different frequency services.}}
\end{figure}

\begin{table}[!t]
\renewcommand{\arraystretch}{1.2}
\caption{Parameters of Frequency Services in Fig.~\ref{fig:f_HD}}
\label{tab:f_para}
\noindent
\centering
    \begin{minipage}{\linewidth} 
    \renewcommand\footnoterule{\vspace*{-5pt}} 
    \begin{center}
        \begin{tabular}{ c || c | c | c | c  }
            \toprule
             &$\boldsymbol{H}$ &$\boldsymbol{D}$ &$\boldsymbol{R}$  & $\boldsymbol{\Delta P_{L0}}$  \\
             &$\boldsymbol{\mathrm{[MWs/Hz]}}$ &$\boldsymbol{\mathrm{[MW/Hz]}}$ &$\boldsymbol{\mathrm{[MW]}}$  & $\boldsymbol{\mathrm{[MW]}}$ \\
            \cline{1-5}
            $\boldsymbol{\mathrm{Case\,II}}$ & $3.2172$  & $0.0169$  & $2.7609$ & $1.5$  \\
            \cline{1-5}
            $\boldsymbol{\mathrm{Case\,IV}}$ & $1.1998$  & $0.9939$  & $1.3586$ & $1.5$\\
           \bottomrule
        \end{tabular}
        \end{center}
    \end{minipage}
    \vspace{-0.25cm}
\end{table}

\begin{table}[!b]
\renewcommand{\arraystretch}{1.2}
\caption{Computational Performance of Different Cases}
\label{tab:time}
\noindent
\centering
    \begin{minipage}{\linewidth} 
    \renewcommand\footnoterule{\vspace*{-5pt}} 
    \begin{center}
        \begin{tabular}{ c || c | c | c | c  }
            \toprule
             &$\boldsymbol{\mathrm{Case\,I}}$ &$\boldsymbol{\mathrm{Case\,II}}$ &$\boldsymbol{\mathrm{Case\,III}}$  & $\boldsymbol{\mathrm{Case\,IV}}$   \\
            \cline{1-5}
            $\boldsymbol{\mathrm{Time\,[s/step]}}$ & \multirow{1}{*}{$24.79$} & \multirow{1}{*}{$49.58$} & \multirow{1}{*}{$45.61$} & \multirow{1}{*}{$60.79$}  \\
           \bottomrule
        \end{tabular}
        \end{center}
    \end{minipage}
    \vspace{-0.35cm}
\end{table} 

To further reveal the reason behind this improvement, time-domain simulating is carried out for Case~II and IV, with the results depicted in Fig.~\ref{fig:f_HD} and the corresponding parameters in Table~\ref{tab:f_para}. Although all the frequency security constraints can be maintained in both cases, different trends in frequency and RoCoF evolution are shown in the figure. In Case II, a large amount of total inertia ($3.22\,\mathrm{MWs/Hz}$) is observed, which results in a small initial RoCoF ($-0.22\,\mathrm{Hz/s}$) and relatively large settling time. It should also be noted that in order to maintain the frequency nadir constraint, redundant frequency response ($R$) from SGs is dispatched during steady-state, and hence only the nadir constraint is binding. Note that the small amount of damping in this case is due to the load-dependent damping. In Case IV, however, two different frequency services can be coordinated to achieve optimal performance in terms of operation cost and frequency trajectories. On one hand, less inertia is needed ($1.20\,\mathrm{MWs/Hz}$) due to the virtual damping provision, which yields a fast response and less oscillation. On the other hand, with virtual inertia and damping being decision variables, both the frequency nadir and steady-state constraints are binding, which means the frequency services are utilized more efficiently. 

The computational time in different cases is also illustrated in Table~\ref{tab:time}. Case I shows the shortest computational time since no virtual inertia or damping is provided and the optimization problem is relatively easy to solve. The computational time increases in Case II and III where only virtual inertia ($49.58\,\mathrm{s/step}$) or damping ($45.61\,\mathrm{s/step}$) is allowed. Although more time ($60.79\,\mathrm{s/step}$) is needed for the proposed method, it does not increase significantly, compared with Case II and III, and is still within an acceptable range. 


\color{black}

\color{black}

\subsection{Impact of Non-essential Load Shedding delay}\label{sec:5.3}
\begin{figure}[!t]
    \centering
    \vspace{-0.35cm}
	\scalebox{1.2}{\includegraphics[]{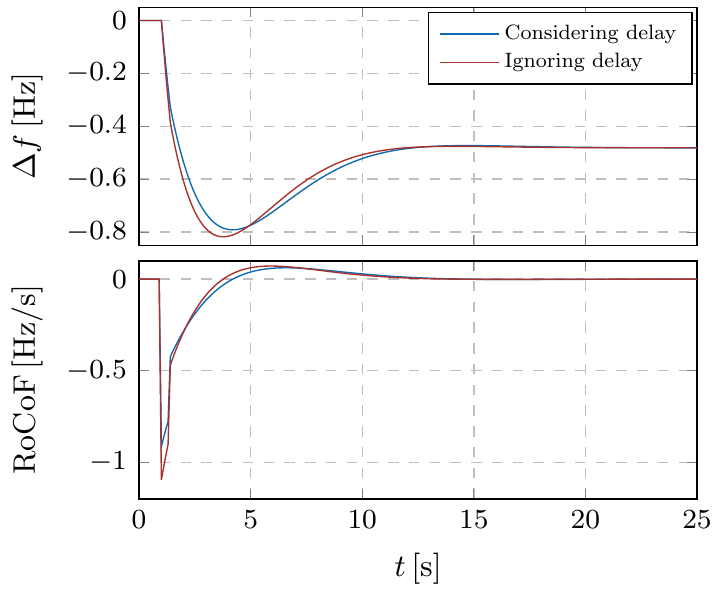}}
    \caption{\label{fig:f_delay}{Microgrid frequency and RoCoF trajectories after an islanding event with and without delays.}}
    \vspace{-0.35cm}
\end{figure}
During an unintentional islanding event, the non-essential load shedding could effectively reduce the disturbance size due to the loss of generation from the maingrid. However, the associated time delay, if being neglected or not properly modeled may lead to an underestimation of frequency variations and RoCoF, threatening the microgrid frequency stability. Therefore, it is necessary to understand the impact of this time delay on the microgrid frequency security. 

Microgrid scheduling problems are performed twice where the time delay of non-essential load shedding is ignored and considered respectively. The frequency regulation performance of the two cases is studied by feeding the obtained solutions at the same hour in the two cases to the time domain simulation. The dynamic response of frequency deviation and system RoCoF in two different cases are depicted in Fig.~\ref{fig:f_delay}. It is clear from the figure that if the time delay due to the non-essential load shedding is ignored when formulating the microgrid frequency security constraints (blue curves), both the frequency nadir and maximum RoCoF exceed the limits during an islanding event with magnitudes being $-0.818\,\mathrm{Hz}$ and $-1.096\,\mathrm{Hz/s}$ respectively. This is because less inertia and damping are prepared for the frequency response during the microgrid scheduling process, compared with the amount they are actually needed. Hence, ignoring non-essential load shedding may lead to frequency constraint violations thus endangering the microgrid security and reliability, which indicates the necessity of appropriately modeling this time delay when developing the microgrid frequency security constraints. On the contrary, once the time delay is included in the frequency security constraints, both the frequency nadir and maximum RoCoF are kept within the permissible range, as shown by the red curves in Fig.~\ref{fig:f_delay}, demonstrating the effectiveness of the proposed model. 

\begin{figure}[!t]
    \centering
	\scalebox{1.2}{\includegraphics[]{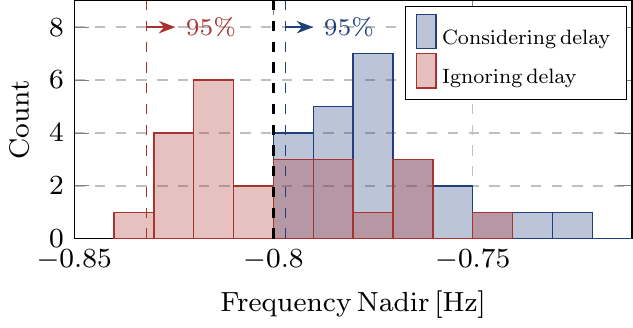}}
    \caption{\label{fig:His_nadir2}{{Frequency nadir histogram (not) considering the time delay of non-essential load shedding.}}}
    \vspace{-0.35cm}
\end{figure}

Moreover, the frequency nadir histogram of 24 hours is also plotted with/without the time delay of non-essential load shedding being considered. As indicated by the red bars in Fig.~\ref{fig:His_nadir2}, the frequency nadir exceeds the limits ($-0.8\,\mathrm{Hz}$) for more than 50\% of the time when the time delay of non-essential load shedding is ignored, { } If an unintentional islanding event occurs during those hours, the microgrid frequency security cannot be guaranteed based on the frequency services scheduled in this way. This unconservative estimation of the frequency nadir is solved by the proposed method where the delay due to the non-essential load shedding is explicitly modeled. Represented by the blue bars, the frequency nadir of all the 24 hours in this case can be effectively maintained above the limit.

\begin{figure}[!t]
    \centering
    \vspace{-0.35cm}
	\scalebox{1.2}{\includegraphics[]{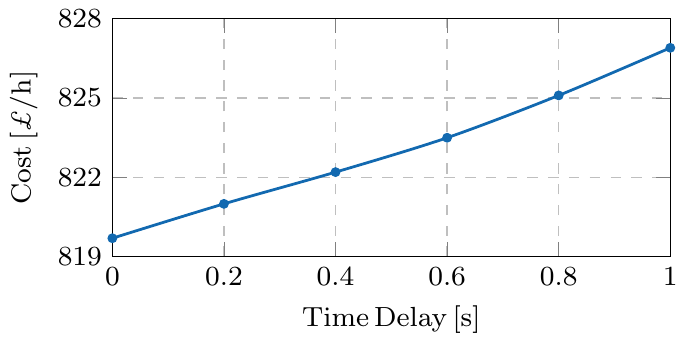}}
    \caption{\label{fig:Cost_delay}{Microgrid operation cost with different non-essential time delay.}}
    \vspace{-0.55cm}
\end{figure}

The influence of non-essential load shedding delay varying from 0 to 1s \cite{4497040} on the microgrid operation cost is also studied. As shown in Fig.~\ref{fig:Cost_delay}, a larger time delay increases the microgrid operation cost in an approximately linear fashion. This is due to the fact that a larger time delay makes the load shedding become less effective in supporting the post-contingency frequency evolution. However, the overall impact on the operation cost is insignificant, which indicates that microgrid frequency security can be maintained by the proposed method with slightly increased cost when considering non-essential load shedding delay. 

\subsection{Value of Dynamically Optimizing Virtual Inertia and Damping}
\begin{figure}[!b]
    \centering
    \vspace{-0.35cm}
	\scalebox{0.95}{\includegraphics[]{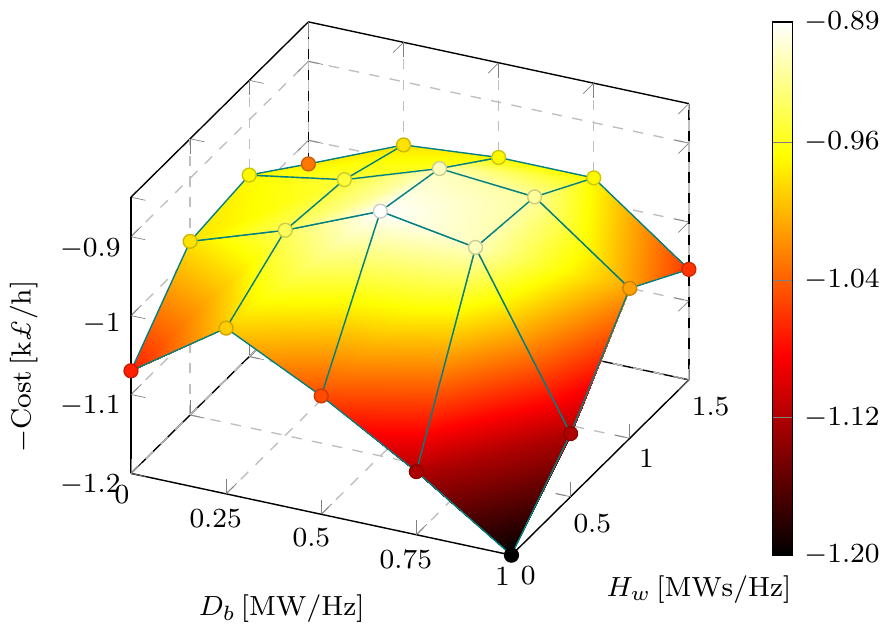}}
    \caption{\label{fig:FixedHD}{Microgrid operation cost with fixed virtual inertia and damping provision.}}
    \vspace{-0.35cm}
\end{figure}
The economic value of dynamically optimizing the control parameters of IBRs in software-defined microgrids to provide the best frequency support performance is demonstrated by comparing it with the conventional fixed-level provision of virtual inertia and damping. The averaged microgrid operation cost at different combinations of virtual inertia and damping is depicted in Fig.~\ref{fig:FixedHD}. Note that for ease of visualization, the $z$-axis is defined as minus cost and next, the impact of $D_b$ and $H_w$ is analyzed respectively. It is understandable that as the amount of virtual damping increases, on one hand, it facilitates the frequency dynamics thus reducing the operational cost (absolute value), and on the other hand, more power needs to be reserved in storage systems, which limits their effectiveness as energy buffers during normal operation. The latter dominates the former at a higher amount of virtual damping, hence resulting in an overall decreasing then increasing trend as shown in the figure. Similarly, the virtual inertia provision from WTs, helps to maintain the frequency constraints, but leads to underproduction due to the recovery effect \cite{9066910}. Therefore, a cost increment can be observed at a higher virtual inertia level. The lowest operation cost ($\mathrm{0.89\,k\pounds/h}$) is attained at $D_b=0.5\,\mathrm{MW/Hz}$ and $H_w=0.5\,\mathrm{MWs/Hz}$, which is still higher than the cost of the proposed approach ($\mathrm{0.82\,k\pounds/h}$). Moreover, this `optimal' fixed combination of virtual inertia and damping would vary depending on system conditions and cannot be tuned straightforwardly in the real world. Nevertheless, the proposed framework that dynamically optimizes virtual inertia and damping in software-defined microgrids outperforms the conventional strategy where the fixed control parameters can only be designed offline.

\begin{figure}[!t]
    \centering
	\scalebox{1.2}{\includegraphics[]{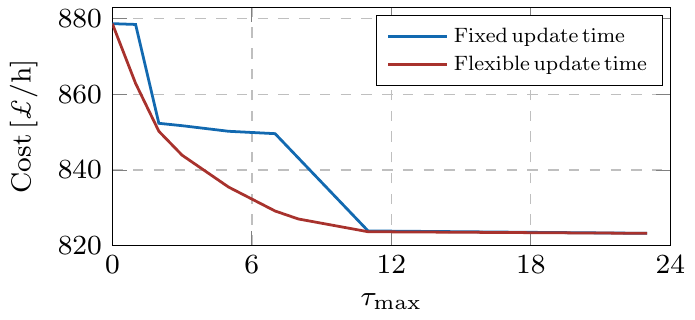}}
    \caption{\label{fig:Update_Freq}{Microgrid operation cost with different $\tau_{\mathrm{max}}$.}}
\end{figure}
\color{black}
On the other hand, the microgrid operator and IBR owner may not want to update the IBR control parameter too frequently due to the security and resilience concerns as discussed in Section~\ref{sec:4.2.3}. Therefore, in order to investigate the impact of maximum allowable IBR control parameter update frequency, the microgrid operation cost with various $\tau_{\mathrm{max}}$ is depicted in Fig.~\ref{fig:Update_Freq}, where two different cases are considered. The ``fixed update time" refers to the case where the IBR control parameters can only be updated every $24/(\tau_{\mathrm{max}}+1)$ hours, whereas in the case with ``flexible update time", when to change the IBR control parameters is determined by the optimization. It can be observed that the microgrid operation cost decreases as $\tau_{\mathrm{max}}$ increases in both cases, since larger IBR control parameter update frequency leads to more frequency control flexibility, to account for the time-varying operating conditions. Moreover, for the blue curve, it cannot fully utilize this flexibility as when to update the IBR control parameters is fixed, instead of being optimized as for the red curve, thus presenting higher cost at smaller $\tau_{\mathrm{max}}$. It can be further spotted that the cost reduction as $\tau_{\mathrm{max}}$ increases from 11 to 23 is negligible, which indicates that it may not be necessary to update the IBR control parameter at each time step of the scheduling, to achieve economic and resilient operation. 

\begin{figure}[!b]
    \centering
	\scalebox{1.2}{\includegraphics[]{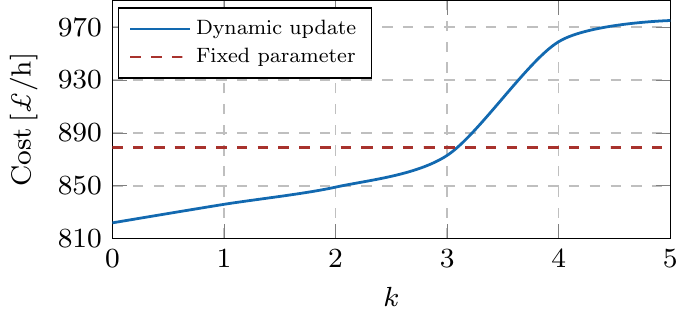}}
    \caption{\label{fig:Robust}{Microgrid operation cost with different degrees of robustness against IBR control parameter update failure.}}
\end{figure}

\textcolor{black}{The optimal value of the control parameter update frequency could then be any value between 0 and this threshold, depending on the risk of communication failure and cyber-attacks. If this risk is small, the optimal frequency tends to the threshold. On the contrary, if the risk is large, the optimal update frequency approaches 0. Theoretically, this optimal control parameter update frequency can be determined by formulating an optimization problem, where the cost reduction as a function of the control parameter update frequency and the cost increment as a function of control parameter update failure should be determined. Although the exact derivation of those functions can be problematic, some approximations can be made based on sensitivity analysis, which will be tackled in future work.}


\subsection{Impact of Robust Frequency Constraints Against Parameter Update Failure}
\begin{figure}[!t]
    \centering
	\scalebox{1.2}{\includegraphics[]{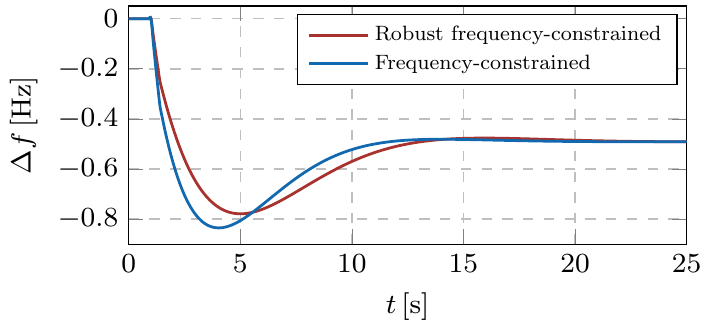}}
    \caption{\label{fig:f_robust}{Post-event microgrid frequency trajectories under (robust) frequency-constrained formulation.}}
\end{figure}

Robust frequency constraints are formulated in Section~\ref{sec:3.3}, to ensure the frequency security considering the potential IBR control parameter update failure. The influence of the maximum number of potential failure ($k$), is investigated with the total number of controllable IBR being 10 (5 wind and 5 storage units). The results are demonstrated in Fig.~\ref{fig:Robust} with two situations being considered. The blue curve represents the proposed method, where the control parameters can be dynamically updated at each time step of the scheduling, whereas the red curve is included as a reference where the control parameter is fixed, thus not suffering from the parameter update failure. It can be observed from the figure that as the number of potential failure increases, the microgrid operation cost rises since more frequency support needs to be preserved. Furthermore, after a certain point ($k>3$), the cost induced by the robust operation becomes even higher than the case with fixed parameters, which means that if the potential failure rate of IBR control parameter update is high, it is not worthwhile for the microgrid operator to dynamically optimize the IBR control parameter compared with the fixed parameter operating strategy. It also highlights the importance of a reliable and accurate communication system in software-defined microgrids. \textcolor{black}{Note that in the case of a parameter update failure, the control parameters are set to zero. Alternatively, the control parameters can be set to their previous value, to improve the overall performance, since more frequency responses are available from the IBRs. However, due to the changes in system operating conditions, the feasibility of the IBRs providing the required frequency responses cannot be guaranteed, thus not considered in this work.} \textcolor{black}{Moreover, the frequency response in the case of $k=1$ is depicted in Fig.~\ref{fig:f_robust}. It can be observed that with one IBR control parameter update failure, the robust frequency-constrained formulation is able to maintain frequency constraints, whilst the conventional approach fails, thus illustrating the effectiveness of the proposed method.}

\section{{Conclusion}} \label{sec:6}
This paper presents a novel software-defined microgrid scheduling model for optimal frequency regulation, to ensure frequency security subsequent to an unintentional island event. Within the framework of software-defined microgrid, dynamic optimization of virtual inertia and damping from IBRs and non-essential load shedding are utilized to facilitate the post event frequency dynamics, while considering the side effects of these services, namely, the IBR control parameter update failure and the time delay of non-essential load shedding. The effectiveness of the proposed method is validated through case studies in IEEE 33-bus system, which illustrates the economic value of simultaneous and dynamic optimization of virtual inertia and damping from IBRs. It is also demonstrated that the overlook of the time delay of non-essential load shedding and the potential IBR control parameter update failure would lead to over-optimistic results.


\vspace{-0.25cm}
\bibliographystyle{IEEEtran}
\bibliography{bibliography}

\vfill
\begin{IEEEbiography}[{\includegraphics[trim={1.5cm 1.0cm 1.5cm 5.0cm}, width=1in,clip,keepaspectratio]{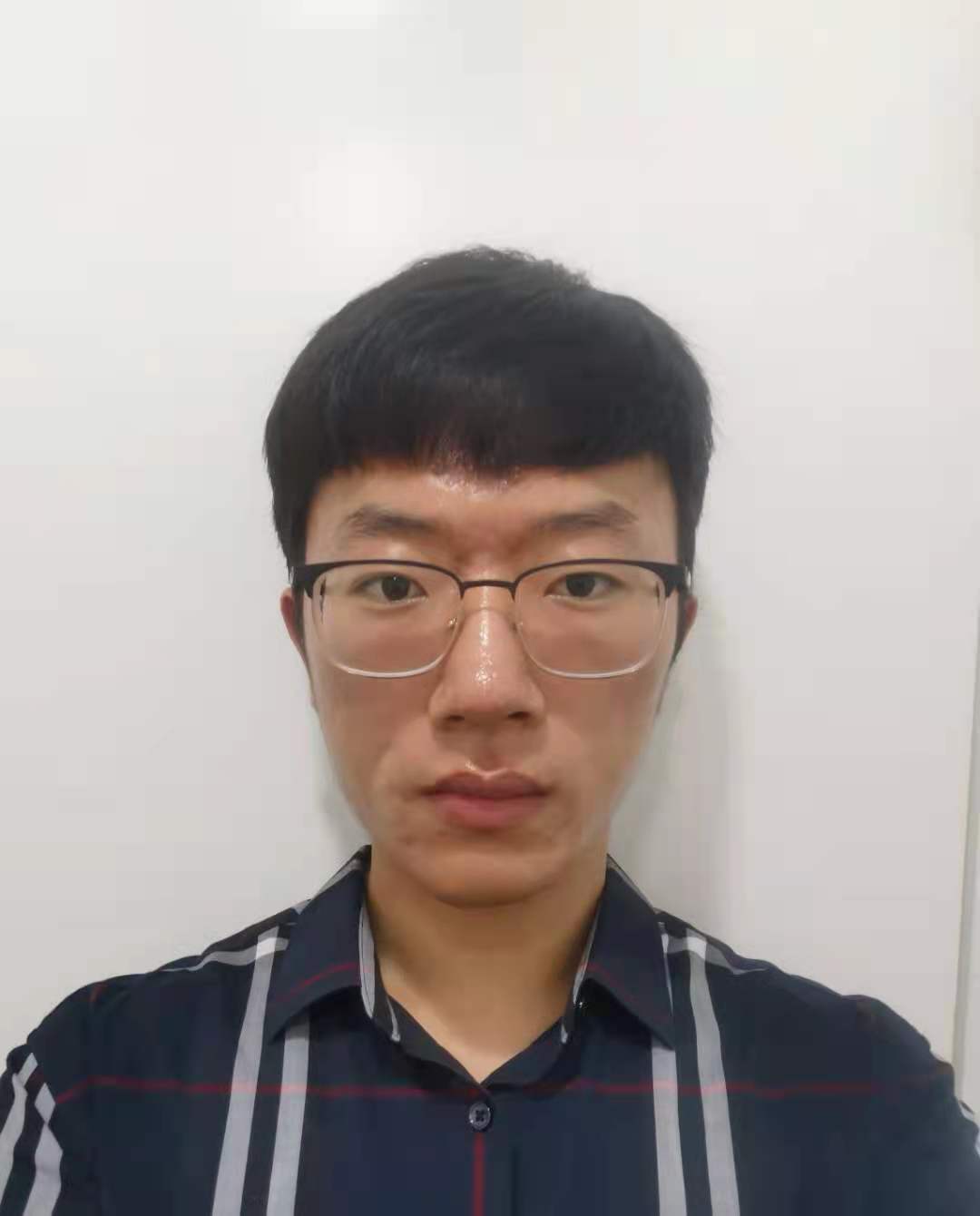}}]{Zhongda Chu}
(S'18-M'22) received the M.Sc. degree in electrical engineering and information technology from the Swiss Federal Institute of Technology, Zürich, Switzerland, in 2018 and the Ph.D. degree in electrical engineering from Imperial College London, London, U.K., in 2022. He is currently a research associate with the Department of Electrical and Electronic Engineering, Imperial College London. His research interests include control and optimization of power systems with high power electronics penetration. 
\end{IEEEbiography}

\begin{IEEEbiography}[{\includegraphics[trim={0cm 2.0cm 0cm 0.0cm}, width=1in,clip,keepaspectratio]{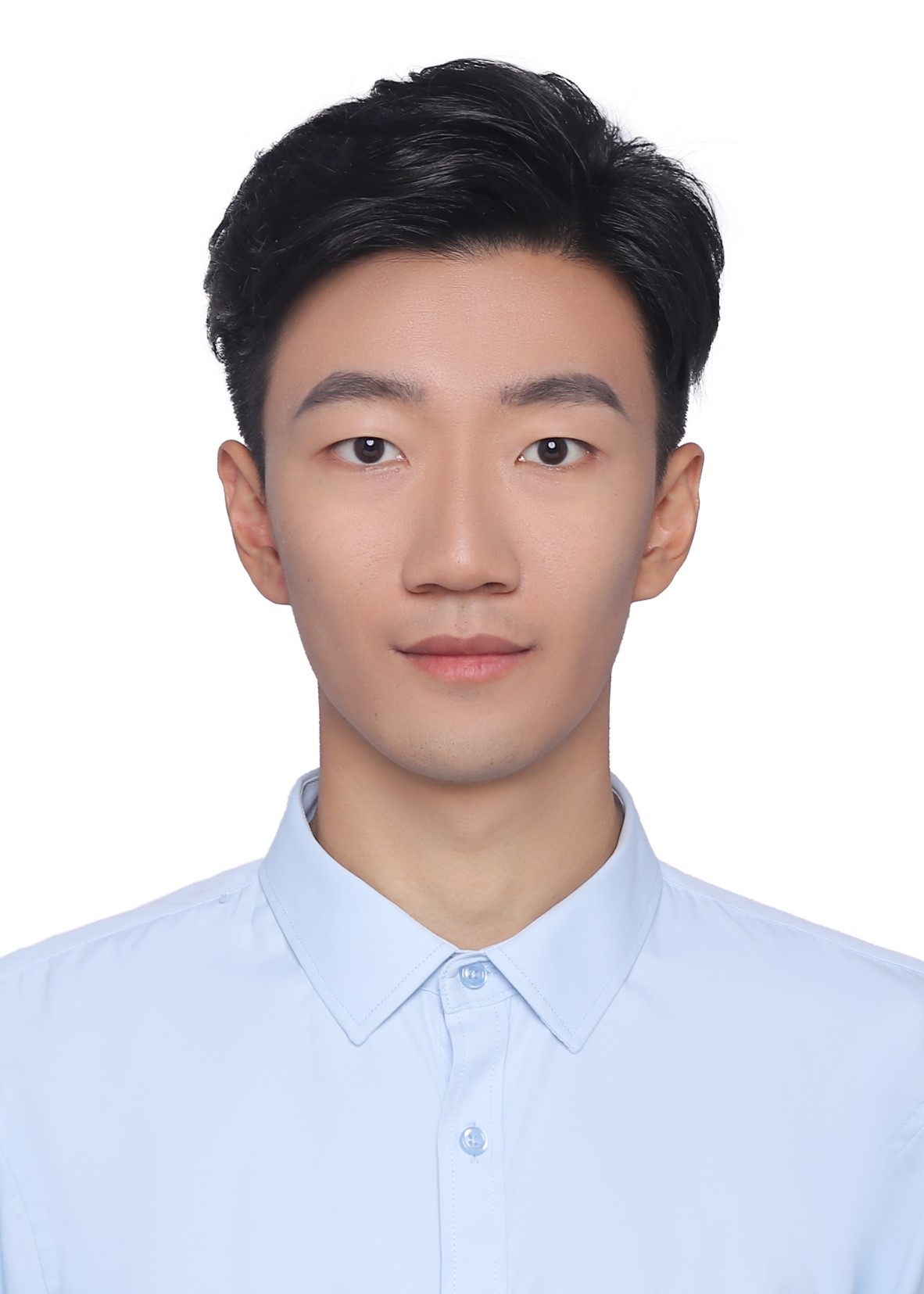}}]{Guoxuan Cui}
(S'23) received the B.Eng. degree in electrical and electronic engineering from North China Electric Power University (Beijing) and the University of Bath in 2020, and the M.Sc. degree in future power networks from Imperial College London in 2021. He is currently pursuing a Ph.D. in the Control and Power Group at the Department of Electrical and Electronic Engineering, Imperial College London. His research focuses on grid-forming converter penetrated power system optimization and synthetic inertia forecasting.
\end{IEEEbiography}

\begin{IEEEbiography}[{\includegraphics[trim={.1cm 1.2cm .1cm .1cm}, width=1in,clip,keepaspectratio]{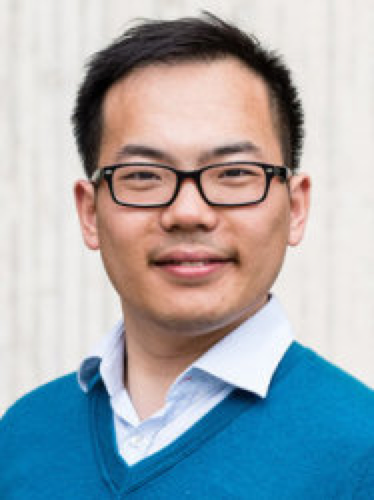}}]{Fei Teng}
(M'15–SM’21) received a B.Eng. degree in Electrical Engineering from Beihang University, China, in 2009, and M.Sc. and Ph.D. degrees in Electrical Engineering from Imperial College London, U.K., in 2010 and 2015, respectively, where he is currently a Senior Lecturer with the Department of Electrical and Electronic Engineering. His research focuses on the power system operation with high penetration of Inverter-Based Resources and the cyber-resilient and privacy-preserving cyber-physical power grid.
\end{IEEEbiography}

\end{document}